\documentclass{plainstyle}

\usepackage{hyperref}

\journal{Journal of Logical and Algebraic Methods in Programming}

\usepackage{fontenc}[T1]
\usepackage{amsmath,amssymb,amsthm,stmaryrd}
\usepackage{hyperref}
\usepackage{color}
\usepackage{times}
\usepackage{enumerate}
\usepackage[usenames,dvipsnames,svgnames,table]{xcolor}

\newcommand{\ilam}[1]{}
\newcommand{\mar}[1]{}
\newcommand{\note}[1]{}

\newcommand{\ilacs}[1]{}

\newcommand{\jpc}[1]{}

\newcommand{\ma}[1]{#1}

\newcommand{\ilaifthenelse}[3]{\ensuremath{\ke{if}\ \, #1\ \ke{then}\ #2\ \ke{else}\ \,#3}}
\newcommand{\true}{\val{true}}

\newenvironment{myenumerate}
               {\begin{enumerate}\vspace{-7pt}
               \topsep0pt\parskip0pt\partopsep0pt\itemsep0pt\leftmargin0pt\itemsep2pt\labelwidth0pt\labelsep3pt}
               {\vspace{-4pt}
               \end{enumerate}}
  \newenvironment{myitemize}
               {\begin{itemize}\vspace{-5pt}\topsep0pt\parskip0pt\partopsep0pt\itemsep0pt\leftmargin0pt\itemsep2pt\labelwidth0pt\labelsep3pt}
               {\end{itemize}}

\newcommand{\lolli}{\multimap}
\newcommand{\with}{\mathbin{\binampersand}}
\newcommand{\tensor}{\otimes}
\newcommand{\one}{\mathbf{1}}
\newcommand{\bang}{{!}}

\def\ptp#1{\pa{#1}}%
\def\coco{\ensuremath{\text{CO}_{2}}} %

\newcommand{\code}[1]{\co{#1}}
\newcommand{\qmark}{{{\textup{\texttt{\symbol{`\?}}}}}}
\newcommand{\atom}[2][]{\code{#2}}
\newcommand{\atomIn}[2][]{\atom[#1]{#2}{\qmark}}
\newcommand{\atomOut}[2][]{\atom[#1]{#2}{\bang}}

\newcommand{\sot}[1]{\texttt{#1}} 
\newcommand{\st}[1]{\texttt{#1} }
\newcommand{\co}[1]{\texttt{#1}} 
\newcommand{\cl}[1]{\textsc{#1\,}} 
\newcommand{\val}[1]{\ensuremath{\mathsf{#1}}} 
\newcommand{\var}[1]{\textit{#1}} 
\newcommand{\ch}[1]{\ensuremath{\mathsf{#1}}} 
\newcommand{\pa}[1]{\ensuremath{\mathsf{#1} }}
\newcommand{\pr}[1]{\textit{#1}} 
\newcommand{\lvr}[1]{\textit{#1}} 
\newcommand{\ke}[1]{{\mbox{\texttt{\textbf{#1}}}}}

\begin{document}
\begin{frontmatter}

\title{Combining behavioural types with security analysis}

\author[1]{Massimo Bartoletti}
\author[2]{Ilaria Castellani} 
\author[3]{Pierre-Malo Deni\'elou}
\author[4]{Mariangiola Dezani-Ciancaglini}
\author[5]{Silvia Ghilezan}
\author[5]{Jovanka Pantovic}
\author[6]{Jorge A. P\'{e}rez}
\author[7]{Peter Thiemann}
\author[8]{Bernardo Toninho}
\author[9]{Hugo Torres Vieira}  

\address[1]{Dipartimento di Matematica e Informatica, University of Cagliari, Italy}
\address[2]{INRIA Sophia Antipolis, France}
\address[3]{Royal Holloway, University of London, UK\footnote{Now at
    Google Inc.}}
\address[4]{Dipartimento di Informatica, Universit\`a di Torino, Italy}
\address[5]{Faculty of Technical Sciences, University of Novi Sad, Serbia}
\address[6]{University of Groningen, The Netherlands}
\address[7]{University of Freiburg, Germany}
\address[8]{Department of Computing, Imperial College London, UK}
\address[9]{IMT Institute for Advanced Studies Lucca, Italy}

\begin{abstract}
  Today's software systems are highly distributed and interconnected,
  and they increasingly rely on communication to achieve their goals;
  due to their societal importance, security and trustworthiness are
  crucial aspects for the correctness of these systems. Behavioural
  types, which extend data types by describing also the structured
  behaviour of programs, are a widely studied approach to the
  enforcement of correctness properties in communicating systems.
  This paper offers a unified overview of proposals based on
  behavioural types which are aimed at the analysis of security
  properties.
\end{abstract}

\end{frontmatter}




\section{Introduction}

Computing systems are omnipresent nowadays; besides their classical
application domains, they have entered into multiple dimensions of our
lives, from business to leisure, from finance to e-government and
health, from global logistics to home appliances, to mention but a
few.  Most of these systems are distributed over the network, and thus
rely heavily on \emph{communication} to carry out their tasks; for
example, in the financial world where decisions are taken in the
global market, or in the context of emerging home appliances that
autonomously shop for groceries on our behalf.  Given their importance
and societal impact, it is crucial that these communicating systems
behave in a reliable way.  This is not an easy task, since they have
to run over \emph{open networks}, where they can be targeted by
malicious parties trying to threaten their functionality, or to seize
or compromise sensitive data.  It is therefore fundamental to develop
rigorous (and scalable) techniques to ensure the reliability and
security of these systems.

Distributed systems are very challenging to analyse, for a variety of
reasons: these range from their intrinsic heterogeneous nature, to the
possible presence of untrusted components, to the complexity of the
interactions and of their induced behaviours.  In the realm of
programming languages, \emph{type systems} represent a
well-established technique to ensure program properties.  Types allow
programmers to single out programs that are correct (i.e., error-free,
for a certain class of errors) at compile-time, just by inspecting
their source code.  Examples of errors that may be excluded by type
systems are the inability of an object to handle a method call
(message not understood), {\em races} (competition among concurrent
programs for some shared resource), which may lead to inconsistent
states or unexpected behaviours, and \emph{communication errors},
caused by non-matching expectations of two communicating partners.

In today's open and highly distributed computing environment, security
flaws of various kinds may arise, and it is crucial to exclude them
before programs are deployed.  Static and dynamic techniques for
ensuring access control and secure information flow were originally
conceived for operating systems.  In the last two decades, spurred by
the pioneering work of~\cite{DenningD77} and~\cite{volpanojcs96} on
static analysis for secure information flow, type systems targeting
security properties have been gradually introduced both into
specification languages such as process calculi
~\cite{Bossi-Focardi-Piazza-Rossi'04,Crafa-Rossi'05,Focardi-Gorrieri'01,Hennessy'05,Hennessy-Riely'02,Honda-Vasc-el'00,Honda-Yoshida'07,Kobayashi'05,Pottier'02,Ryan-Schneider'99}
and into full fledged programming
languages~\cite{SPARK,jif,HedinBBS14,SantosR14,flowcaml}.

Classical data types are an abstract specification of {\em what}
programs compute (i.e., the outcome of computations).  By contrast,
\emph{behavioural types} (\emph{BTs}) specify also {\em how} programs
compute (i.e., the structure of computations) thus giving a more
intensional description of their behaviour (the reader is referred
to~\cite{wg1soar} for an extensive discussion on this point).  BTs are
particularly suited in dealing with concurrent and communicating
processes, whose semantics is based on their {\em interactive
  behaviour}.  In this case, BTs may be viewed as abstract protocols,
describing the causal and branching structure of communications among
a number of parties. First introduced in the setting of process
calculi~\cite{DBLP:conf/concur/Honda93,DBLP:conf/esop/HondaVK98},
behavioural types for communication-centric systems have been studied
for a variety of calculi and languages since the late nineties.  For a
recent survey, the reader is referred to~\cite{wg3soar}.

The design of behaviourally typed languages requires some care: on the
one hand, one would like to keep a simple and intuitive programming
flavour (to ensure that typed abstractions may be widely translated
into practice); on the other hand, typed languages must be developed
bearing in mind the complexity (not to mention decidability) of their
associated verification techniques.  Hence, BT-based approaches must
be fine-tuned to keep a balance between expressiveness and feasibility
of analysis.

In this respect, \emph{session
  types}~\cite{DBLP:conf/concur/Honda93,DBLP:conf/esop/HondaVK98}
stand out as a particularly attractive instance of behavioural
types. Session types allow interactions to be structured into basic
units called \emph{sessions}.  Individual interaction patterns are
then abstracted as session types, against which process descriptions
may be checked.  The expressiveness of session types has enabled their
application in diverse contexts, targeting different programming
models (e.g., functional~\cite{DBLP:journals/tcs/VasconcelosGR06} and
object-oriented
programming~\cite{DBLP:journals/iandc/Dezani-CiancagliniDMY09}), and
addressing also lower-levels of application (namely operating system
design~\cite{DBLP:conf/eurosys/FahndrichAHHHLL06} and middleware
communication protocols~\cite{DBLP:journals/fuin/VallecilloVR06}), to
mention a few.

Although some proposals that promote the use of BTs for the
analysis of security properties have been put forward, the study of
security and trustworthiness properties for typed communicating
systems is still at an early stage. 

Security type systems for variants of the $\pi$-calculus, which
combine security enforcement with other correctness concerns, were
presented in~\cite{Hennessy'05,Hennessy-Riely'02} and
~\cite{Honda-Vasc-el'00,Honda-Yoshida'07}.  A simple type system
ensuring noninterference for the $\pi$-calculus was proposed
in~\cite{Pottier'02}. Subsequently, more refined security type systems
for the $\pi$-calculus were studied in~\cite{Crafa-Rossi'05} and
~\cite{Kobayashi'05}. This last work provides a sophisticated security
analysis, together with a type inference algorithm.

The question of unifying the language and process-based approaches to
security has also been addressed.  The goal was to define translations
from the former to the latter that would preserve both the security
properties and the security types.  A first step in this direction was
taken in~\cite{Honda-Vasc-el'00}, where a typed parallel imperative
language was embedded into a typed $\pi$-calculus. This work was
further pursued in~\cite{Honda-Yoshida'07}, where more powerful
languages, both imperative and functional, were considered.  Among the
types used in these early works, the closest to behavioural types are
the {\em usage types} of ~\cite{Kobayashi'05}, which describe how
processes use their communication channels along computations.  It
should be noted, however, that all these approaches are based on
(simple variants of) the $\pi$-calculus, and not on so-called {\em
  session calculi}, more specialised variants of the $\pi$-calculus
where communication has an explicit logical structure.

Session-typed models focus on open-ended systems, where loosely
coupled parties may synchronise to start a session on a specific
(public) service, thereafter interacting on the private (restricted)
channel of the session.  In general, one expects security properties
to be simpler to enforce within a session than in an open network:
since session participants must conform to their session types, their
behaviour is more disciplined than that of arbitrary, untyped
processes. Indeed, in a trusted session-typed setting, one may focus
on systems where participants communicate according to prescribed
protocols -- this property is often referred to as {\em session
  fidelity} -- and this restricts the range of possible security
flaws. Moreover, within this trusted platform, one may address more
specific security properties, regarding, e.g., the information that is
communicated or the participants who carry out the communications, in
order to achieve a finer tracking of security flaws.  Finally, since
interactions within a session take place on a private channel, session
isolation is guaranteed by construction.

However, this scenario is too restrictive in practice. In realistic
distributed applications, the components are spread across open
networks and cannot be completely controlled. Hence, both the
\emph{closed-network assumption} (ensuring the isolation of session
channels) and \emph{the typed-world assumption} (ensuring the correct
behaviour of all components) have to be lifted.  Only part of the
application may be assumed to be typed, and hence trusted. The
challenge is then to guarantee that the desired security properties
continue to hold when the trusted part interacts with the untrusted
one.

In this document, we present a survey of existing BT-based approaches
aimed at ensuring security properties.  Our review does not only
reflect several different views of secure and trustworthy
communication-based systems, but also highlights how behavioural types
provide a simple conceptual framework to formally approach all such
different visions, from various angles.

The rest of the paper is organised as follows.
Section~\ref{sec:preliminaries} introduces the basic notions of
session types.  In Section~\ref{sec:information flow} we discuss
behavioural types which guarantee access control, secure information
flow, and integrity of data in communication protocols.
Section~\ref{sec:proofcarryingcode} shows how the logical foundation
of session types -- based on the correspondence with linear
logic~\cite{DBLP:conf/concur/CairesP10} -- can be fruitful also for
security. In particular, dependent session types are applied to proof
carrying code and digital certificates.

To some extent, it is fair to say that the approaches described in
Sections~\ref{sec:information flow}--\ref{sec:proofcarryingcode}
consist in extending foundational settings with security concerns or
enhancing existing techniques in order to tackle security issues. In a
somewhat distinct direction, we find techniques that focus on
guaranteeing that the properties studied in foundational settings can
be transported to more realistic models encompassing open networks:
Section~\ref{sec:protocolabstraction} reviews a compilation of
session-like descriptions into implementations of cryptographic
protocols that ensures that honest session participants are protected
from external interference.  Section~\ref{sec:protocolabstraction}
also describes a theory of contracts that addresses a different notion
of honesty among interacting parties, roughly referring to
participants that behave as promised (contracted) even when engaged in
other interactions with dishonest parties. The protection of
participants from malicious contexts is further addressed in
Section~\ref{sec:gametheory}, by means of game theoretic approaches.

Section~\ref{sec:gradualtyping} presents recent developments,
including two approaches that use dynamic typing, in contrast to the
type systems reviewed in previous sections.  Gradual typing allows
legacy code to be reconciled with varying security policies
(Sections~\ref{gts} and~\ref{gsts}). Section~\ref{ss:adap} describes a
process framework which exploits behavioural types to jointly enforce
run-time adaptation and the combination of access control and secure
information flow. Section~\ref{sec:dynamicdata} reports on a
combination of techniques that address access control with a
behavioural typing system that focuses on role-based protocol
specifications, considering that role impersonation must be duly
\emph{authorized}.  Section~\ref{sec:conclusions} summarises the
various approaches reviewed in the paper, and gives some tracks for
future research.



\section{Preliminaries}
\label{sec:preliminaries}

In this section, we informally present some key notions pertaining to
behavioural types, focusing on session types, which are the kind of
BTs for which most security-oriented extensions have been proposed.
The interested reader is referred to the recent survey~\cite{wg1soar}
for further details on the foundations of behavioural types.

Session types characterise how communication channels are used by
programs. Intuitively, much like the type declaration \var{var}:
\sot{int} tells us that \var{var} will be used to hold integer values,
a type declaration \ch{chan}: \st{SessionType} tells us that \ch{chan}
will be used to hold an access point to a channel that will be used by
the program according to \st{SessionType}. However, \st{SessionType}
may actually refer to several stages of the usage of \ch{chan}. A
simple instance of a session type is
\[\st{S}_1 = \st{!}\sot{int}.\st{?}\sot{bool}.\st{end}\]
Then, the type declaration $\ch{chan}: \st{S}_1$
indicates that \ch{chan} is first used to output an integer and then
to input a boolean.  More precisely, `$\st{!}$' denotes output,
`$\st{?}$' represents input, `$\st{.}$' captures sequentiality, and
`$\st{end}$' denotes no further usage.

A fundamental feature of session types is that they capture
\emph{linear} interactions.  Linearity amounts to forbidding competing
communications on the same channel, what is sometimes referred to as a
{\em communication race}.  Session types exclude communication races
but they allow for the specification of alternative behaviours
(controlled by one party), described via so-called \emph{branching}
and \emph{selection} types.  Branching represents an {\em external
  choice}, to be solved by the communicating partner, while selection
represents an {\em internal choice}, typically placed in the branch of
a conditional.  Hence, alternatives are modelled by a ``menu'' and a
corresponding choice: the former is usually implemented via a set of
available inputs, and typed using branching; the latter is usually
implemented via an output, and typed using selection.  Alternatives
are conveniently identified by \emph{labels}.  For example, the
session type
\begin{center}
$\st{S}_2=\,\cl{number} \st{?}\sot{int}\st{.!}\sot{int}\st{.end + } \cl{condition}\st{?}\sot{bool}\st{.end}$
\end{center} 
may be used to describe a process that is waiting for the choice of
either one of its offered options, identified by labels \cl{number}
and \cl{condition}.  This process can be safely composed with, e.g., a
process willing to first choose the \cl{number} option and then to
send and receive an integer. 

It is worth noticing that since types talk about the current stage of
the protocol, type preservation, i.e., the property that ensures that
well-typedness is preserved under system evolution, holds modulo an
evolution also of types (types are resources that in some sense get
consumed by execution).

Another powerful feature of session types is \emph{(session)
  delegation}, which allows third parties to gain access to an already
established interaction. Realised via channel passing, this mechanism
is useful to specify, e.g., a server that at some stage in the
protocol delegates to a third party the remaining communications with
a client (who will proceed with the protocol, unaware of the exchanges
at the server side).  Carried types can thus be session types
themselves; this way, e.g., the session type
\[\st{S}_3 = \st{?}(\st{!}\sot{int}\st{.end})\st{.end}\]
specifies the reception of a channel, which will be used to output an
integer according to $\st{!}\sot{int}\st{.end}$.

Session types for binary ``client-server'' interaction, as introduced
in~\cite{DBLP:conf/esop/HondaVK98}, capture the two-ended interaction
via the type of one of the endpoints --- which are sometimes
distinguished by so-called
\emph{polarities}~\cite{DBLP:journals/acta/GayH05}, denoted $+$ and
$-$: this way, e.g., \ch{chan}$^+$ and \ch{chan}$^-$ would denote the
two endpoints of \ch{chan}. So, if one endpoint is used according to a
session type, it is immediate to recover the characterisation of the
other endpoint by replacing inputs by outputs, branching by selection
and conversely --- a notion usually referred to as \emph{duality}.
The dual of a session type $\st{S}$ is denoted $\overline{\st{S}}$.
This way, e.g., the duals of session types $\st{S}_1$, $\st{S}_2$ and
$\st{S}_3$ above are respectively:
\[\begin{array}{l}
\overline{\st{S}_1} = \st{?}\sot{int}\st{.!}\sot{bool}\st{.end}\\
\overline{\st{S}_2} = \cl{number}\st{!}\sot{int}\st{.?}\sot{int}\st{.end
$\oplus$ }\cl{condition}\st{!}\sot{bool}\st{.end} \\
\overline{\st{S}_3}=\st{!}(\st{!}\sot{int}\st{.end})\st{.end}\\
\end{array}\]
More recently, following the approach introduced in~\cite{CHY08},
generalisations of session types that capture multiparty interactions
have been put forward. In such a setting of {\em multiparty sessions}, 
pairwise linear interaction is still ensured,
via the use of dedicated intra-session channels (or session
indexes~\cite{BCDDDY08}).
In this case, the session types of individual parties no longer
suffice to capture the entire structured interaction: global
specifications, called \emph{global types}, have been
introduced to specify sequencing information for message exchanges among the various
communicating parties. Since a global type explicitly
identifies the involved parties, it is possible to extract from it the
individual contributions of each party via a \emph{projection}
function --- these contributions are commonly referred to as \emph{local types}. 

In the above examples and in the sequel the syntactic categories are
distinguished by using different fonts, namely: \sot{sort types,
  session types, contracts}, \cl{choice labels}, \val{values,
  participants}, \var{variables, processes, terms}, \ke{keywords}.



\section{Security Types for Communication-Centred Calculi}\label{sec:information flow}

An increasingly relevant security issue is that of preserving the {\em
  confidentiality} of private data that is hosted on cloud
infrastructures and/or manipulated by Web services and
applications. Protection of data confidentiality requires two
complementary techniques: {\em access control}, which restricts the
access to the original data, allowing only trusted users to read them,
and {\em secure information flow}, which prevents the propagation of
legally accessed data to untrusted users, thus ensuring end-to-end
confidentiality. Compared to access control, secure information flow
may be viewed as additionally restricting the access to transferred or
transformed data, when these have been computed using sensitive data.
Type systems for access control and secure information flow are
reviewed in Sections~\ref{ss:accon} and~\ref{ss:secif}, respectively.

Another important security property is data {\em integrity}, which is
often presented as the dual of confidentiality and may be similarly
expressed as a combination of access control and secure information
flow. While confidentiality requires that data should not be released
to untrusted destinations, integrity requires that data should not be
affected by untrusted parties.  Type systems for data integrity are
discussed in Section~\ref{icd}.

\subsection{Access Control}\label{ss:accon}
An early work featuring behavioural types for access control is
\cite{lapadulafsen07}, which presents a type system for COWS (Calculus
for Orchestration of Web Services)~\cite{PT12}, a formalism for
specifying and combining services, as well as modelling their dynamic
behaviour.  The COWS language provides a primitive for killing
processes, possibly provoking the abortion of ongoing sessions. This
type system allows for the specification and enforcement of {\em
  policies} for regulating the exchange of data among services.  To
implement such policies, programmers can annotate data with sets of
participants authorised to use and exchange these data.  The typed
operational semantics uses these annotations to guarantee that
computations proceed correctly.

For example, consider a standard buyer-seller-bank protocol, in which:
\begin{myenumerate}
\item the buyer
asks the seller for some item and receives back a price;
\item the buyer may either accept the price and send a credit card
  number to the bank, or turn down the offer.
\end{myenumerate}
In this scenario, the desired policy is that the credit card number
should be accessible only to the buyer and the bank, but not to the
seller.  Therefore, the type system of \cite{lapadulafsen07} validates
processes implementing the protocol described above, but not variants
of it in which by mistake the buyer would send the credit card number
to the seller. This policy is represented by decorating the data
\val{creditCardNumber} with participants \pa{agent} and \pa{bank}:
$\{$\val{creditCardNumber}$\}_{\pa{agent}, \pa{bank}}$.

\medskip

Another related service-oriented calculus is SCC (Service Centered
Calculus)~\cite{BBCNLLMMRSVZ06}. The work~\cite{kolundzijawsfm08}
enriches a variant of SCC, investigated in~\cite{BM08}, with security
levels for controlling access rights. In the original calculus,
communications may either follow fixed protocols or use pipelines. A
\emph{pipe constructor} $\pr{P}< \var{x}< \pr{Q}$, similar to that of
the Orc language~\cite{Orc2009}, replaces the variable $\var{x}$ in (a
freshly spawned copy of) the process $\pr{Q}$ with a value
sent by the process $\pr{P}$.  In the new calculus, processes are {\em
framed} \cite{PSS05} by security levels.  A process framed by a level
$\ell$ can exercise rights of security level not exceeding
$\ell$. Security levels are assigned to service definitions, clients
and data. In order to invoke a service, a client must be endowed with
an appropriate clearance, and once the service and client agree on the
security level, the data exchanged in the initiated session will not
exceed this level. The calculus of \cite{kolundzijawsfm08} comes
equipped with a behavioural type system that statically ensures these
security properties.

In the buyer-seller-bank protocol described above, the protection of
the credit card number is ensured by giving it a security level which
is incomparable with the level of the seller, but is smaller or
equal to the levels of the buyer and the bank. More formally, if the
processes representing the seller, the buyer and the bank are framed
by the security levels $\ell_1$, $\ell_2$ and $\ell_3$, respectively,
then it is enough to assign to the credit card number the type
$\sot{nat}$ with some level $\ell$ such that $\ell\not\leq\ell_1$,
$\ell\leq\ell_2$ and $\ell\leq\ell_3$.

\medskip

The work~\cite{capecchiiandc13} adopts the same treatment of access
control as~\cite{kolundzijawsfm08}, ascribing security levels to both
participants and data, but it gains flexibility thanks to the
mechanism of {\em delegation}. Namely, it allows participants to get
around access control restrictions by delegating the handling of
sensitive data to other participants with higher credentials. For
instance, instead of sending a bank connection to the buyer, thus
explicitly involving the bank in the interaction, the seller may
delegate to the bank the part of the interaction that deals with the
credit card in a way that is transparent to the client. The type
system that ensures access control has an explicit type constructor to
track delegation; it allows the delegated part of a session type to be
marked.

\medskip

Notably, all the types investigated
in~\cite{lapadulafsen07},~\cite{kolundzijawsfm08},
and~\cite{capecchiiandc13} are behavioural types. While those used
in~\cite{capecchiiandc13} are standard session types enhanced with
security constraints, the others are more general instances of BTs. A
comparison between the three access control approaches described above
is not easy, since the underlying calculi offer different interaction
patterns, namely a primitive for session killing
in~\cite{lapadulafsen07}, a pipeline constructor
in~\cite{kolundzijawsfm08} and channel delegation in
\cite{capecchiiandc13}.  In our view, the most challenging construct
for data protection is delegation, since it allows a transparent
change of ownership for a given communication channel.

\subsection{Secure Information Flow}\label{ss:secif}

As already mentioned above, the work~\cite{capecchiiandc13} considers
a calculus for multiparty sessions with delegation, enriched with
security levels for both participants and data, and equipped with an
access control mechanism.  This work also defines a secure information
flow property, formalising the preservation of data confidentiality.
Finally, a session type system is proposed, which introduces secure
information flow requirements in the typing rules, in order to
simultaneously ensure the noninterference property and the standard
behavioural properties prescribed by session types. Such
security-enhanced session types are an instance of behavioural types
specifying both the sequencing of communication actions and the
constraints between their security levels.  The
study~\cite{capecchiiandc13} revealed an interesting interplay between
the constraints used in security types and those used in session types
to ensure properties like communication fidelity and progress. In
essence, session-typed processes are less prone to security
flaws. This point will be discussed in some depth later in this
section. We now describe the approach of~\cite{capecchiiandc13} in
more detail.

The {\em secure information flow} property defined in
\cite{capecchiiandc13} is based on the observation of messages while
they are being exchanged. As usual in the secure information flow
literature, the observation power depends on the level of the
observer. An observer of level $\ell$ can only see messages of
security level lower than or equal to $\ell$.  For simplicity, we
assume here just two security levels $\top$ and $\bot$ (although
\cite{capecchiiandc13} deals with a general lattice of security
levels).  A message or I/O communication action whose carried value is
of security level $\top$ (respectively, $\bot$) will be called
``secret'' or ``high'' (respectively, ``public'' or ``low'').

We shall use here a simple syntax where $\ch{c}(\var{x}^\ell).\pr{P}$
denotes the input on channel $\ch{c}$ of a value of level $\ell$, to
be replaced for $\var{x}^\ell$ in the continuation process $\pr{P}$,
and $\ch{c}<\val{v}^\ell>. \pr{P}$ denotes the output on channel
$\ch{c}$ of the value $\val{v}$ of level $\ell$, to be followed by
$\pr{P}$. For simplicity, we shall only deal with binary sessions
here, so $\ch{c}$ will range over binary session endpoints of the form
$\ch{s}^+, \ch{s}^-$, where $\ch{s}$ is the name of some established
session.

Secure information flow is usually formalised via the notion of {\em
  noninterference}~\cite{GoguenMeseguer1982}. Noninterference
essentially means that low outputs should not depend on high inputs.
Then, a typical insecure information flow, also called {\em
  information leak}, arises when different high inputs cause different
low messages to be exchanged, as in the following process, where we
assume $\var{x}^\top$ to be a boolean variable:
\[\ilaifthenelse{\var{x}^\top}{\ch{s}^+! <\val{1}^\bot>}{ \ch{s}^+! <\val{2}^\bot>}
\]

 Another source of information leaking is the possible blocking of a high
input action, in the case where the environment does not offer the expected high
message (which is something one cannot control, since two environments
differing only for the presence or absence of a high message cannot be
distinguished by a low observer).  For instance, the process:
\[\ch{s}^+?(\var{x}^\top).\ch{s}^+!<\val{1}^\bot>\] emits the low
output ``\val{1}'' only if it first receives the high input from the
environment. Assuming again $\var{x}^\top$ to be a boolean variable,
the session type of channel $\ch{s}^+$ in the above process is
$\st{?}\sot{bool}\st{.!}\sot{int}\st{.end}$. On the other hand, this
process cannot be typed in the security-enhanced session type system,
since it exhibits a ``level drop'' from the input to the output.

The subsequent paper~\cite{CCD15} moves one step further by equipping
the above calculus with a monitored semantics, which blocks the
execution of processes as soon as they attempt to leak information.
This monitored semantics induces a safety property: a process is safe
if none of its computations is blocked by the monitored
semantics. This property is called {\em information flow safety}, and
it is proved to strictly imply the noninterference property
of~\cite{capecchiiandc13}.  This is expected, since information flow
safety requires the successful execution of all individual
computations, while noninterference is a property of the set of
computations of a process, which may hold even if some of the
computations exhibit information leaks.

The approach in~\cite{capecchiiandc13,CCD15} may be summarised as
proposing three increasingly precise means for tracking information
leaks in sessions: a syntactic property (typability), a local semantic
property (safety), and a global semantic property (security).

We illustrate the difference between typability, safety and security
in~\cite{capecchiiandc13,CCD15} by means of a simple example that
should convey the appropriate intuitions.  In this example, typability
requires the absence of any ``level drop'' from the expression tested
by a conditional to a subsequent communication, while safety requires
the same condition but only in computations that may actually occur.

Consider a conditional whose $\top$-level condition is true and whose
\ke{then} branch sends $\bot$-level data, while its \ke{else} branch
sends $\top$-level data (whose value does not really matter, since
this branch is never taken):
\[ \ilaifthenelse{\true^\top}{\ch{s}^+! <\val{1}^\bot>}{\ch{s}^+! <\val{2}^\top>}
\]
This process is secure because it always exhibits the same public
behaviour, but it is neither safe nor typable.  Consider now a variant
of the above process, where the two branches of the conditional are
swapped:
\[ \ilaifthenelse{\true^\top}{\ch{s}^+! <\val{2}^\top>}{\ch{s}^+! <\val{1}^\bot>}
\]
This process is still not typable, but it is now both safe and secure,
since the $\ke{else}$ branch is never taken and thus the level drop
cannot occur in any computation. An example of a typable process is:
\[ \ilaifthenelse{\true^\top}{\ch{s}^+! <\val{1}^\top>}{\ch{s}^+! <\val{2}^\top>}
\]
In this process, channel $\ch{s}^+$ has the security-enhanced type
$\st{!}\sot{int}^\top\st{.end}$.

\subsubsection*{Discussion}

There appears to be an influence of classical session
types~\cite{CHY08} upon security types~\cite{volpanojcs96}.  Indeed,
one of the causes of insecure information flow in a concurrency
scenario is the possibility of different \emph{termination behaviours}
in the branches of a high conditional (i.e., a conditional which tests
a high expression). This may give rise to the so-called
\emph{termination leaks}.  In session calculi, there are three
possible termination behaviours: proper termination, deadlock and
divergence. Then, a termination leak may occur, for instance, if one
branch of a high conditional terminates while the other diverges or
deadlocks, assuming successful termination is made explicit by an
observable action.  Session types help to contain this phenomenon, by
imposing some uniformity in the termination behaviours of conditional
branches: for instance, a terminating branch cannot coexist with a
diverging branch, as exemplified below. They also prevent local
deadlocks (due to communication errors within a session) as well as
some global deadlocks, thus limiting the possible sources of abnormal
termination. Note that one may also use typing to ensure (proper)
termination. For instance, termination is ensured by the session type
system studied in~\cite{DBLP:conf/concur/CairesP10}, as shown
in~\cite{Perez2014}. The system in~\cite{DBLP:conf/concur/CairesP10}
is the basis for the dependent session types reviewed in
Section~\ref{sec:proofcarryingcode}.

Since the two branches of a conditional must have the same session
types for all channels, we cannot for example type the process:
\[\ilaifthenelse{\var{x}^\top}{\ch{s}^+! <\val{1}^\top>}{\ke{rec}\; \pr{X}. \ch{s}^+! <\val{2}^\top>.\pr{X}}
\]
which could cause a termination leak. Typing also prevents termination
leaks due to bad matchings of data, like in the process:
\[\ilaifthenelse{\var{x}^\top}{\ch{s}^+! <\val{1}^\top>}{ \ch{s}^+! <\var{x}^\top\val{+3}>}
\]
where we assume that $\var{x}^\top$ is replaced by a boolean value.
However, the security-enhanced typing considered
in~\cite{capecchiiandc13,CCD15} does not prevent {\em global
  deadlocks} due to bad matchings of protocols in interleaved
sessions, like in the process:
\[\begin{array}{l}\ke{if }{\var{x}^\top}\ke{ then }{\ch{s}^+?(\var{y}).\ch{s}^+! <\val{1}^\top>.\ch{r}^+?(\var{z}).\ch{r}^+!<\val{2}^\top>}\\
\phantom{\ke{if }{\var{x}^\top}}\ke{ else }{\ch{s}^+?(\var{y}).\ch{r}^+?(\var{z}).\ch{s}^+! <\val{1}^\top>.\ch{r}^+!<\val{2}^\top>}\\
| ~\ch{s}^-! <\val{3}^\top>.\ch{s}^-?(\var{t}).\ch{r}^-!<\val{4}^\top>.\ch{r}^-?(\var{u})
\end{array}\]
Here, if the \ke{then} branch is taken the process will terminate
successfully, while if the \ke{else} branch is taken the interaction
will deadlock. The security-enhanced session type of the channels
$\ch{s}^+$, $\ch{r}^+$ is
$\st{?}\sot{int}^\top\st{.!}\sot{int}^\top\st{.end}$, while the
channels $\ch{s}^-$, $\ch{r}^-$ have the dual session type
$\st{!}\sot{int}^\top\st{.?}\sot{int}^\top\st{.end}$.
These global deadlocks are forbidden by the more refined behavioural
type systems in \cite{BCDDDY08,K02,P13}.


\subsection{Integrity of Communicated Data}\label{icd}
We now turn to the issue of data integrity, as investigated in the
work~\cite{bonellijfp05}.  We start by considering the standard
User-ATM-Bank example~\cite{DBLP:conf/esop/HondaVK98}. In response to
a deposit request by the user, a malicious ATM could send to the bank
an amount of money that is different from that communicated by the
user, consequently altering the balance obtained from the bank. This
change is transparent to the typing, since it does not modify the
communication protocol.  This means that the following processes,
where channels $\ch{s},\ch{r}$ are used respectively for the
interaction between the user and the ATM and between the ATM and the
bank:
\[\begin{array}{lll}
\pa{user}&=& \ch{s}^+ ! <\val{userId}>. \ch{s}^+ ! <\val{depositAmount}>\\
\pa{ATM}&=& \ch{s}^- ? (\var{userId}). \ch{s}^- ? (\var{depositAmount}).\\
&& \ch{r}^+ ! <\var{userId}>. \ch{r}^+ ! <\var{depositAmount}\val{-10}>\\
\pa{bank}&=& \ch{r}^- ? (\var{userId}). \ch{r}^- ? (\var{depositAmount})
\end{array}\]
can be typed since channels $\ch{s}^+$, $\ch{r}^+$ have type
$\st{!}\sot{string}\st{.!}\sot{int}\st{.end}$, while the channels
$\ch{s}^-$, $\ch{r}^-$ have the dual session type
$\st{?}\sot{string}\st{.?}\sot{int}\st{.end}$.

In order to cope with this kind of misbehaviour, {\em correspondence
  assertions} \cite{GJ03} are incorporated in the theory of session
types~\cite{bonellijfp05}.  Two correspondence assertions can be
paired by the keywords $\ke{begin}$, $\ke{end}$ and their values allow
the integrity of the communicated data to be checked (in this example
\val{userId} and \val{depositAmount}).  The user and the bank
processes with correspondence assertions become:
\[\begin{array}{l}
\pa{user} \,=\, \ke{begin} (\val{userId}, \val{depositAmount}). \ch{s}^+ ! <\val{userId}>. \ch{s}^+ ! <\val{depositAmount}>\\
\pa{bank} \,=\, \ch{r}^- ? (\var{userId}). \ch{r}^- ? (\var{depositAmount}). \ke{end} (\var{userId}, \var{depositAmount})
\end{array}\]
thus allowing the malicious ATM to be discovered,
since the operational semantics requires the same values in paired
correspondence assertions. The session types of the channels remain
unchanged, given that correspondence assertions play the role of
run-time monitoring.
 
\medskip

Compared to standard session types,
session types enhanced with correspondence
assertions may be used to check additional properties, namely:
\begin{myitemize}
\item the source of information,
\item whether data are propagated as specified across multiple parties,
\item if there are unspecified communications between parties, and
\item if {\em the data being exchanged have been modified} in some
  unexpected way.
\end{myitemize}

More recently, \cite{BMV12} presents a $\pi$-calculus with assume and
assert operations, typed using a session discipline that incorporates
refinement formulae written in a fragment of Multiplicative Linear
Logic~\cite{Girard87}. This original combination of session and
refinement types, together with the well-established benefits of
linearity, allows very fine-grained specifications of communication
protocols in which refinement formulae are treated as logical
resources rather than persistent truths.
 
 Another related paper is~\cite{DBLP:conf/concur/BocchiHTY10}, where
session types for multiparty sessions are enriched with constraints on
the content of the exchanged messages, conditions on the choice of
sub-conversations to follow, and invariants on recursion.



\section{Logical Approaches to Security based on Behavioural Types}
\label{sec:proofcarryingcode}

As already discussed, session types consist of high-level
specifications of the communication behaviour of distributed,
concurrent processes along channels. Historically, these
specifications capture input/output behaviour, replication (or
persistency), branching and selection behaviours, and recursion;
enabling static verification of protocol compliance (or session
fidelity).
However, classic session types are not expressive enough to describe
properties of data exchanged in communications, nor to certify such
properties in a distributed setting, where the user of a service does
not have access to the application source code.
Both issues are a fundamental problem in today's world, given the
increasing pervasiveness and complexity of distributed services, for
which simple descriptions of communication behaviour are insufficient
characterisations of the rich, high-level contracts these services are
intended to follow.

To address the issue of lack of expressiveness in terms of properties
that can be characterised by session types, extensions to the session
framework have been presented (e.g., the work~\cite{bonellijfp05},
already discussed in section~\ref{icd},
and~\cite{DBLP:journals/jfp/SwamyCFSBY13}).  Recently, logical
foundations for session types have been established via Curry-Howard
correspondences with linear logic \cite{DBLP:conf/concur/CairesP10}.
Besides clarifying and unifying concepts in session types, such
logical underpinnings provide natural means for generalisation and
extensions. One such extension to \emph{dependent session types}
allows for expressing and enforcing complex properties of data
transmitted during sessions \cite{toninho11ppdp}. This is achieved by
interpreting the first order quantifiers of intuitionistic linear
logic as input and output constructs, in which it is possible to refer
to the actual value that is communicated in the types themselves. By
combining this with a data language that is itself dependently typed
(e.g., in the style of LF \cite{Harper:1993:FDL:138027.138060}), we
are able to specify arbitrary properties of the communicated data in
such a way that the {\em proof objects} that witness the desired
properties are themselves exchanged during communication. Moreover,
the solid logical foundations of the approach enable further
(logically grounded) extensions to the data language to capture
features of interest in an almost immediate way, such as digital proof
certificates and proof object erasure through proof irrelevance and
affirmation modalities \cite{DBLP:conf/cpp/PfenningCT11}.

Sections \ref{sec:lldst}, \ref{ss:pirrel}, and \ref{ss:affirm}
overview extensions of session types with (value) dependent types,
proof irrelevance, and affirmation, respectively.

\subsection{Linear Logic and Dependent Session Types}\label{sec:lldst}

Linear logic is a logic of resources and evolving state, where
propositions can be seen as resources that interact with each other
and evolve (i.e., change state) over time. These are the fundamental
characteristics that allow for the development of the Curry-Howard
correspondence between linear logic and session types.

The work of \cite{DBLP:conf/concur/CairesP10} interprets the
propositional connectives of linear logic as the session types
assigned to $\pi$-calculus channels in such a way that linear logic
proofs can be interpreted as typing derivations for $\pi$-calculus
processes. Moreover, the computational procedure of proof
simplification or \emph{proof reduction} is directly mapped to
inter-process communication, thus obtaining a true correspondence
between the dynamics of proofs and the dynamics of communicating
processes. The connectives of linear logic are linear implication
$\st{A} \lolli \st{B}$, which is interpreted as the input session type
(i.e., input a session channel of type $\st A$ and continue as $\st
B$); its dual, multiplicative conjunction $\st{A} \tensor \st{B}$,
which is naturally interpreted as session output (i.e., output a
session channel of type $\st{A}$ and continue as $\st B$); the
multiplicative unit, $\one$, denoting the inactive or terminated
session; additive conjunction $\st{A} \with \st{B}$ denoting an offer
of a choice, meaning that a session of type $\st{A} \with \st{B}$ will
be able to offer along the session channel either $\st A$ or $\st B$,
the choice of which is left to the session client; dually, additive
disjunction $\st A \oplus \st B$ denotes alternative behaviour, and so
a session of type $\st A\oplus \st B$ will unilaterally choose to
behave as either $\st A$ or $\st B$. Finally, the linear logic
exponential $\bang \st A$ is mapped to replication, in which a session
of type $\bang \st A$ will offer a potentially unbounded number of
instances of the behaviour $\st A$.
Moreover, a fundamental aspect of proof theory is proof composition,
also known as a \emph{cut}. In the interpretation, cuts are mapped to
\emph{process composition}; two processes using disjoint sets of
resources interact along a fresh session channel, where one offers a
session and the other uses it to produce some other session behaviour.

Recently, \cite{toninho11ppdp} extended this framework of
propositional linear logic as session types to incorporate
\emph{dependent} session types by moving to a first-order setting,
introducing the two quantifiers $\forall \lvr x{:}\tau . \st A$ and
$\exists \lvr x{:}\tau .\st A$, where $\lvr x$ may occur free in $\st
A$. The quantification variable is itself typed, with a domain of
quantification $\tau$. The language of terms inhabiting $\tau$ is a
typed $\lambda$-calculus, which is left as general as possible, with
the usual soundness requirements of progress, substitution and type
preservation. The interpretation of these session types is (typed)
term output for the existential $\exists \lvr x{:}\tau . \st A$ and
term input for the universal $\forall \lvr x{:}\tau. \st A$. Thus, a
session of type $\exists \lvr x{:}\tau .\st A$ outputs a term $\pr M$
of type $\tau$ and proceeds as type $\st A\{\pr M/\lvr x\}$, whilst a
session of type $\forall \lvr x{:}\tau . \st A$ behaves in a dual
manner.

By making the quantification domain dependently typed, the authors
obtain a session type system where processes exchange data but also
proof objects that can denote properties of said data. For instance,
the type:
\begin{equation}
\st{UpInterfaceP}(\lvr x)\triangleq  \lvr x : \forall \lvr n{:}\sot{int}. \, \forall
\lvr p{:} (\lvr n > \val 0). \,
\exists \lvr y{:}\sot{int}. \, \exists \lvr q{:} (\lvr y>\val 0) .\, \one
\label{eq:pc}
\end{equation}
denotes a session that will input an integer $\val n$\footnote{The
  different font for integers ($\lvr n$ and $\val n$) is due to our
  convention that distinguishes between variables and values.} and a
\emph{proof} that $\val n$ is greater than $\val 0$, and will then
output back an integer $\val y$, itself greater than $\val 0$, and a
proof of this fact. Well-typedness ensures that these properties hold
at run time due to the existence of these proof objects, making this
dependently-typed session framework a \emph{de facto} model of
proof-carrying code.

\subsection{Proof Irrelevance}\label{ss:pirrel}

In a distributed setting, the proof-carrying framework above requires
not only that proof objects exist during type-checking, but also
enforces that they are transmitted at run time. However, it is often
the case that we want the specified properties to hold but we do not
want to exchange the proof objects.  Indeed, omitting the
communication of proof objects may be sensible when the communicating
parties have established trust by some external means, or when the
properties are easily decidable and the proof objects can be
\emph{synthesised} by a decision procedure.  For instance, in the
example above, it is straightforward to check that the communicated
numbers are indeed strictly positive.

To model the possibility of omitting proofs at run time, the work of
\cite{DBLP:conf/cpp/PfenningCT11,toninho11ppdp} extends the framework
by internalising in the proof object language the concept of
\emph{proof irrelevance} \cite{Pfenning:2001:IEP:871816.871845},
through a modality denoted $[\tau]$, which types terms of type $\tau$
that can be \emph{safely} erased at run time. This notion of erasure
safety essentially means that such terms can never be used to compute
values that are not themselves erasable. For instance, the type above
can be rewritten as:
\begin{equation}
\st{UpInterfaceI}(\lvr x)\triangleq  \lvr x : \forall \lvr n{:}\sot{int}.\, \forall
\lvr p{:} [\lvr n > \val 0]. \,
\exists \lvr y{:}\sot{int}. \, \exists \lvr q{:} [\lvr y>\val 0] .\, \one
\label{eq:pi}
\end{equation}
remarking the fact that the proof objects $\lvr p$ and $\lvr q$ must
be present for type-checking purposes, but they are not used in a
computationally significant fashion at run time and therefore can be
safely omitted. This process of erasing proofs at run time is done in
two steps: first all instances of proof irrelevant types and terms are
replaced with the unit type and element (denoted $\sot{unit}$ and
$\langle \rangle$, respectively). Since this procedure does not remove the
communication step where the proof objects were previously exchanged,
we may exploit the type isomorphisms,
$$
\begin{array}{c}
\forall \lvr x{:}\sot{unit}.\st A \cong \st A\\
\exists \lvr x{:}\sot{unit}.\st A \cong \st A
\end{array}
$$
to consistently remove the communication overhead.  An alternative
technique familiar from type theories is to replace sequences of data
communications by a single communication of pairs. When proof objects
are involved, these become $\Sigma$-types ({sum types}) which are
inhabited by pairs. For example, we can rewrite $\st{UpInterfaceI}$ as:
$$
\st{UpInterfaceI}_2(\lvr x)\triangleq  \lvr x : \forall \lvr p{:}(\Sigma \lvr n{:}\sot{int}. 
[\lvr n > \val 0]). 
\exists \lvr q{:}(\Sigma \lvr y{:}\sot{int}. [\lvr y>\val 0]) .\one
$$
This solution is popular in type theory, where $\Sigma \lvr
x{:}\tau.[\sigma]$ is a formulation of a {\em subset type}, $\{\lvr
x{:}\tau \mid \sigma \}$. Conversely, bracket types $[\sigma]$ can be
written as $\{\lvr x{:}\sot{unit} \mid \sigma \}$, except the proof
object is always erased. Under some restrictions on $\sigma$ (i.e.,
decidability of the underlying theory), subset types can be seen as
predicate-based type refinements.

\subsection{Affirmation and Digital Certificates}\label{ss:affirm}
The examples above showcase what can be seen as two extremes in a
spectrum of \emph{trust}.  In the type given in \eqref{eq:pc} no trust
between the parties is assumed and therefore all proof objects must be
made explicit in communication at run time.  On the other hand, proof
irrelevance as represented by the type given in \eqref{eq:pi} models a
scenario of \emph{full trust}, where no proof objects are expected at
run time. In practice, there are tradeoffs between trust and fully
explicit proofs. For instance, when downloading a large application we
may be willing to trust its safety if it is digitally signed by a
reputable third party, but if we are downloading and running a piece
of Javascript code embedded in a web page, we may insist on an
explicit proof that it adheres to our security policy. To make these
tradeoffs explicit in session types, \cite{DBLP:conf/cpp/PfenningCT11}
also incorporates in the framework a notion of \emph{affirmation}
(from modal logic) of propositions and proofs by principals. Such
affirmations can be realised through explicit digital signatures on
proofs by principals, based on some underlying public key
infrastructure.

The key component to model these certificates is the addition of a
type $\Diamond_{\pa K} \tau$ to the framework, which types objects
that assert the property $\tau$, signed by principal $\pa K$ using its
private key.  An affirmation object is built by taking the original
proof object that asserts $\tau$ and signing it accordingly.
Superficially, this may seem redundant insofar as the certificate
contains the proof object itself. However, checking a digitally signed
certificate may be much faster than checking the validity of a proof,
so we may speed up the system if we simply trust $\pa K$'s
signature. Moreover, when combining certificates with proof
irrelevance, we may construct certificates where parts of the original
proof object have been erased, and so we have in general no way of
reconstructing the original proofs. In these cases we necessarily
trust the signing principal $\pa K$ to accept $\tau$ as true.

Combining affirmation and proof irrelevance it is possible to
model the following,
$$
\st{fpt} {:} \forall \lvr f{:}\sot{nat} \rightarrow \sot{nat} . \forall \lvr p {:}
\Diamond_{\pa{verif}}[\Pi \lvr x{:}\sot{nat}. \lvr{f(x)} \leq \lvr x]. \exists
\lvr y{:}\sot{nat}.\exists \lvr q{:} [\lvr{y=f(y)}].\one
$$
which expresses the type of a service that first inputs a function
$\lvr f$, then accepts a verifier's word that each natural number is a
prefixed point of $\lvr f$, and finally returns a fixed point of $\lvr
f$ to its client.  Observe that object $\lvr p$ is a
\emph{certificate} of the fact that $\lvr f$ satisfies this property.
In realistic scenarios, such as proof-carrying file systems
\cite{Garg10oakland}, the use of affirmation and proof irrelevance
results in substantially less communication overheads when compared to
proof-carrying code in the sense of~\cite{neculapopl97}, where the
proof objects become too big to be transmitted and checked every time
a file is accessed.



\section{Secure Interactions with Untrusted Components} 
\label{sec:protocolabstraction}

Session type systems are able to provide some safety and liveness
guarantees for a whole distributed system, as long as all participants
are well typed and the network is trusted.  In many realistic
settings, however, these assumptions do not hold.

A first approach towards a more realistic scenario is to consider an
untrusted network.  The solution, currently used in every-day life, is
to perform session communications over secure channels, such as those
provided by the Transport Layer Security (TLS) protocol. This ensures
that well-typed participants will interact safely (precisely, as
safely as the TLS protocol allows) within an untrusted environment.

A second, more general scenario is when some of the (multiparty)
session participants are not trusted to be well typed, i.e., they are
not trusted to respect the session specification.  This covers those
cases in which, for instance, participants rely on implementations
provided by non-reliable third parties, or when they may be controlled
by an adversary.  In some cases, not respecting the communication
pattern (e.g., skipping mandatory messages, not respecting branching)
is indeed a security issue. The questions are then the following: what
properties can still be ensured for compliant participants? Which
cryptography should be used to protect the session? How to ensure that
all compliant participants share an identical view of a session
execution?

\subsection{A Secure Protocol Compiler}
\label{sec:secure-protocol-compiler} 

The works~\cite{corinD07,corincsf07,corinjcs08} offer a first answer
to these questions (the journal paper~\cite{corinjcs08} merges and
extends both \cite{corincsf07} and \cite{corinD07}).  The proposed
language, expressed as a type language with a global graph-like
representation (called session graphs), includes messages, roles, and
sessions; it does not support parallelism or asynchrony.  A secure
implementability condition, defined on session graphs, is identified.

The principle of~\cite{corinjcs08} is to use the session graph
specification to generate a cryptographic protocol (and its
implementation) that will protect the honest participants against any
coalition of compromised peers. The idea is that, in order to ensure
that an incoming message is valid with respect to the session graph
specification, that message should carry enough trustworthy
information to be able to prove that the protocol history was
compliant up to that point. Technically, this is achieved by means of
asymmetric cryptography, using signatures of past messages to convince
the receiver that the specification was followed by all
participants. The minimal (necessary and sufficient) set of signatures
to be transmitted and checked is defined through the notion of
\emph{visibility}. The protocol also relies on other cryptographic
primitives, such as nonces and a cache system, to prevent replay
attacks between session instances or within a given session.

The formal security notion proved in~\cite{corinjcs08} is called {\em
  session integrity}. It says that the messages received and accepted
by all compliant participants are always consistent with correct
projected traces of the session specification.  This approach is
implemented as an extension of OCaml. A compiler has been developed,
which takes as input a session description and produces as output an
OCaml module with a function for each participant.  Any user code
calling one of these functions is guaranteed through the OCaml type
system to statically follow the appropriate local session type. This
is achieved through a monadic programming style. The module's
cryptographic implementation then guarantees that, even in the case of
compromised peers, all the messages seen by uncompromised participants
are consistent with the session specification.  As a case study, the
authors of~\cite{corinjcs08} implement and evaluate a conference
management system with three roles: the program committee, an author,
and a submission manager.

Two different extensions of the approach in~\cite{corinjcs08} are
developed in~\cite{planulconcur09} and~\cite{bhargavancsf09}.  The
work~\cite{planulconcur09} generalises~\cite{corinjcs08} with a
  more abstract setup and a greater session expressiveness. Rather
  than a graph representation, the authors of~\cite{planulconcur09}
  define session specifications with CCS-like processes, which
  represent the desired communication pattern. A history-tracking
  process calculus is used as the lower-level model of a secure
  implementation. The correctness of the history-tracking mechanism
  with respect to the CCS specification is proved using a trace-based
  semantics which adequately models an adversary that can control the
  network and remote peers. This captures and generalises the
  signature-based mechanism of~\cite{corinjcs08}.
The work~\cite{bhargavancsf09} improves~\cite{corinD07,corinjcs08}
with simpler and more efficient cryptography (using a combination of
asymmetric and symmetric cryptography), and extends the session
description language with value annotations (i.e., constraints on
  payload). This extension allows one to model commitments and to
protect the integrity of each payload. As
in~\cite{corinD07,corinjcs08}, a compiler implementation is realised,
which relies on OCaml typing for local protocol conformance, and on a
generated optimised cryptographic protocol implementation for session
integrity.



\subsection{Contract-oriented Service Composition}

In the design of session-typed distributed applications according to a
top-down approach~\cite{CHY08}, a choreography describing the global
interaction behaviour of the application is projected to a set of
local types, which describe the contributions of individual
participants in the application.  Each participant is implemented by a
concrete process: if all these implementations respect their local
types, the overall application is guaranteed to enjoy some correctness
property (e.g., the absence of deadlocks).

In an adversarial setting, however, one cannot assume that the
implementation of an untrusted participant respects its local type;
indeed, participants have full control of the code they run, and they
can even change it at run time.  Any static analysis that requires
inspection of the code of each participant is then pointless:
consequently, properties which are not enforceable by run-time
monitoring (e.g., the absence of deadlocks) cannot be enforced at all
in this adversarial setting.

To cope with this situation, a different design approach has been
proposed where the composition of distributed components is performed
in a bottom-up fashion. %
In this approach, participants first advertise their promised
behaviour as \emph{contracts} to some broker; the broker inspects such
contracts, and creates sessions among participants whose contracts
admit an \emph{agreement}.
For instance, contracts could be binary or multiparty session types,
and agreement could be one of the compliance / compatibility 
relations defined over them~\cite{wg1soar}.
Once these sessions are created, participants can perform the actions
prescribed by their contracts (in the case where they are session
types, this would result in performing the prescribed inputs and
outputs).
An execution monitor can then keep track of the state of each contract
with respect to the participants' actions, which cannot depart from
the actions expected by the advertised contracts; furthermore, the
monitor can establish who is \emph{accountable} at each step, %
i.e.,~responsible for the next interaction. %
Systems developed under this design approach are called
\emph{contract-oriented systems}~\cite{BTZ12sacs}.

While untrusted participants may cause deadlocks in both
contract-oriented and top-down approaches, an advantage of the former
is that one can use contracts to single out the participants which
have breached the agreement, thus causing the deadlock. This can be
associated with sanctions imposed to the culpable participants; these
sanctions could range, e.g.,\ from lowering the participants'
reputation or imposing them a fine, to removing them altogether from
the repository of available services.



\subsection{Honesty in Contract-oriented Systems}

Interacting with ill-typed or untrusted participants may have
non-obvious consequences, especially in the case of applications whose
implementations involve multiple interleaved sessions.  For instance,
consider a simple e-commerce scenario involving three participants: a
seller \ptp{A}, a wholesaler \ptp{B}, and a client \ptp{C} that
interact through binary sessions $\ch s$ (between \ptp{A} and \ptp{B})
and $\ch t$ (between \ptp{A} and \ptp{C}).  Suppose that \ptp{A} is
expected to receive a payment from \ptp{C} (in session $\ch t$), then
order an item to \ptp{B} (in session $\ch s$), wait until the item is
received (still in $\ch s$), and finally ship the item to \ptp{C} (in
$\ch t$).  If the wholesaler \ptp{B} receives the payment but never
sends the item to \ptp{A}, then \ptp{A} becomes unable to ship the
item to \ptp{C}. In turn, \ptp{C} may get stuck and be unable to
advance in other sessions. Clearly, the problem may cascade to affect
other sessions and participants.

A desirable goal for the designer of the seller would be to guarantee
that \ptp{A} is never responsible for some stuck session: %
therefore, even if $\ch s$ is blocked by \ptp{B}, %
participant \ptp{A} will still behave according to his contract in
$\ch t$. %
This property --- called \emph{honesty} --- is formally defined and
investigated in the contract-oriented specification language
\coco~\cite{BTZ12coordination}.  Honesty can be seen as multi-session
well-typedness: %
if a participant is honest, %
his process implementation will behave according to his contracts in
each session he establishes, %
even if other participants will not cooperate. %

In general, a developer would aim at publishing only honest services
that always respect contracts --- even when the other participants are
malicious: otherwise, the service infrastructure may eventually
sanction him for contract breaches.  Since honesty cannot be enforced
by run-time monitoring (it is a sort of deadlock-freedom property),
static analysis techniques for detecting honesty of processes are
required.  While honesty is not decidable in
general~\cite{BTZ12coordination}, %
it can be statically approximated: as usual, the approximation must
stay ``on the safe side'', i.e.,\ if it statically determines that a
service is honest, then this is really the case; otherwise, it may be
either the case that the service is honest or it is not.  In the
literature, analysis techniques for honesty have been proposed using
both type systems~\cite{BSTZ13forte} and model
checking~\cite{BMSZ14maude}.



\renewcommand{\atomIn}[2][]{{\qmark}\atom[#1]{\cl{#2}\!}}
\renewcommand{\atomOut}[2][]{{\bang}\atom[#1]{\cl{#2}\!}}

\subsection{Protection Against Untrusted Brokers} 
\label{sec:gametheory}

In contract-oriented applications, participants advertise their
contracts to some broker, which establishes sessions among
participants whose contracts admit an agreement.
In such a scenario, the agreement property guarantees that --- even in
the presence of malicious participants --- no interaction driven by
the contracts will ever go wrong: in the worst case, if some
participant does not reach his objectives, then some other (dishonest)
participant will be culpable of a contract infringement.

In the above workflow, it is often assumed 
that brokers are trusted, in that they never establish a session in
the absence of an agreement.
In more byzantine scenarios, it may happen that a fraudulent broker
creates a session where participants interact in the absence of an
agreement. %
In this way, the broker may allow an accomplice to swindle an unaware
participant. %
Note that the accomplice may perform his scam while adhering to his
contract, and so he cannot be blamed for violations. %
A crucial problem is how to devise contracts which protect
participants from malicious brokers.  In contexts where brokers are
malicious, contracts should still allow participants to reach their
goals when the other participants are cooperative.  At the same time,
contracts should prevent participants from performing imprudent
actions which could be exploited by malicious participants.

This problem has been addressed
in~\cite{BCPZ15jlamp,DBLP:conf/post/BartolettiCZ13} in a
game-theoretic setting, where session interactions are interpreted as
games over \emph{event structures}
(ESs~\cite{DBLP:conf/ac/Winskel86}), and participants are the players
of these games. %
In this setting, a participant wins in a \emph{play} (a trace of the
ES) when he reaches success, or some other participants can be blamed
for a violation. %
The idea
is that the infrastructure will eventually inflict
sanctions to the participants who have violated their contracts, 
as in~\cite{Mukhija2007qos}.

Two key notions in this model are those of \emph{agreement} and
\emph{protection}.  Agreement is a property of contracts which
guarantees that each honest participant may reach success whenever the
other participants cooperate.  Moreover, if an honest participant does
not reach success, then some other participant can be blamed.  A
contract \emph{protects} its participant if, whenever composed with
any other contract (possibly that of an adversary), the contract
admits at least one \emph{non-losing strategy}, i.e., a strategy that
guarantees that the participant will never end up in a failure state.

The notion of agreement in the game-based model is related
in~\cite{BCPZ15jlamp} to the progress-based notion of compliance in
session types~\cite{Barbanera10ppdp}.  More precisely, two session
types are compliant if and only if, in their interpretation as ESs,
\emph{all} (innocent) strategies are winning.  Hence, compliance
implies agreement, while the converse does not hold. %
This is illustrated by the following example, using the syntax of
Section~\ref{sec:preliminaries}.
Consider the following session types (where labels are to be viewed as branching labels): 
\begin{center}
$\co T_1 = \atomOut{a}.\atomOut{c}.\st{end} \,\oplus\, \atomOut{b}.\st{end}$\qquad
and \qquad
$\co T_2 = \atomIn{a}.\st{end} \, + \, \atomIn{b}.\st{end}$.
\end{center}
where $\co T_1$ is advertised by Player $1$, and $\co T_2$ by Player $2$. %
We have that Player $2$ agrees with the composition of $\co T_1$ and $\co T_2$. %
Indeed, the only innocent strategy for Player $2$ is the one which prescribes him to: 
\begin{itemize}
\item do $\atomIn{a}$ after $\co T_1$ has performed $\atomOut{a}$;
\item do $\atomIn{b}$ after $\co T_1$ has performed $\atomOut{b}$.
\end{itemize}
This strategy is winning, because it leads Player $2$ to the success
state $\st{end}$ in both cases (unless Player $1$ violates $\co
T_1$). %
Similarly, Player $1$ \emph{agrees} with the composition of $\co T_1$
and $\co T_2$: his winning strategy is just to choose the branch
$\atomOut{b}$. %
However, the session types $\co T_1$ and $\co T_2$ are \emph{not}
compliant according to~\cite{Barbanera10ppdp}.  Indeed, if $\co T_1$
takes the internal choice $\atomOut{a}$, then a deadlock state is
reached. %

If brokers are dishonest, 
then they may establish sessions even in the absence of an agreement.
For instance, assume Player $3$ advertises the session type:
\[
  \co T_3 = \atomOut{pay}.\atomIn{receive}.\st{end} \, \oplus \, \atomOut{abort}.\st{end}
\]
A dishonest broker could make Player $3$ interact with another player
with session type: 
\[
  \co T_4 = \atomIn{pay}.\st{end}
\]
Note that Player $3$ does \emph{not} agree with the composition of
$\co T_3$ and $\co T_4$, because if his strategy chooses
$\atomOut{pay}$, then he will lose if the other player does not
perform $\atomOut{receive}$ (as in $\co T_4$), while if it chooses
$\atomOut{abort}$, then he will reach a \emph{tie} state (i.e.,
neither success nor failure). %
However, Player $3$ can protect himself against such dishonest
brokers: a strategy that protects him would be the one which only
chooses $\atomOut{abort}$ (intuitively, doing $\atomOut{pay}$ leads
Player $3$ to lose if the other player never does $\atomOut{receive}$,
while doing $\atomOut{abort}$ leads Player $3$ to a tie state
independently of the contract and of the strategy of the other
player). %

While, in general, it is not always possible to guarantee 
that a set of contracts admit both agreement and protection 
(as proved in~\cite{BCPZ15jlamp}),
it is possible to reconcile these two notions
by relaxing the classical notion of causality, 
i.e.,\ by assuming that some events can occur in the absence
of a causal justification in the \emph{past}, 
provided they have a justification 
in the \emph{future}~\cite{DBLP:conf/post/BartolettiCZ13}.



\section{Emerging Directions}\label{sec:gradualtyping}

While highly expressive fully static type systems can be constructed
and proved sound, not many of them have had an impact on computing
practice.  One reason for this unfortunate under-use is that most
software is not written from scratch, but rather by building on top of
existing components or by modifying them.  Clearly, program
modifications must be written in the ``legacy language'' of the
existing code. Extensions may be connected to the legacy components by
foreign function interfaces or by wrapping the legacy code in web
services and connecting to them via communication channels. In these
cases, the new code can be subject to an expressive type discipline
that enforces structural constraints, but the existing code is used as
is, because it would be too expensive to rewrite existing production
code.

This situation is aggravated in the security setting because security
policies are often stated after the fact, when significant parts of a
system have already been implemented, and they are likely to change in
reaction to newly discovered threats and exploits, or to adhere to new
requirements or contextual conditions.

While legacy code maintenance will always be an issue and integration
is of utmost concern, a more recent trend focuses on anticipating how
a system can react to changes in its external environment or
requirements. Since types may provide specifications of correct
behaviour, adaptation at run time may be guided by type information.

In the remainder of this section we describe some dynamic typing
analysis approaches that cope with partial (static) type information
and type-driven adaptation. We also briefly mention a connection
between techniques for ensuring access control and behavioural typing,
in particular considering the importance of \emph{roles} in
communication-centred systems.

\subsection{Gradual Typing and Security}\label{gts}

One approach to introduce behavioural types into existing systems is
to do it gradually, on a per-module basis, so that typed and less
typed program parts must interoperate. Gradual typing
\cite{SiekTaha2007,SiekTaha2006} addresses exactly this
interoperation.  Gradual type systems have been developed from dynamic
type systems \cite{AbadiCardelliPiercePlotkin1991,Henglein1994}.  They
provide a type \texttt{Dynamic} with operations to \emph{inject} a
value of arbitrary type into \texttt{Dynamic} as well as operations to
\emph{project} a dynamic value to an arbitrary type. While an
injection always succeeds, a projection may fail and throw an
exception if the statically expected type does not agree with the
run-time type of the dynamic value. For example, the injection for
type \texttt{int} maps $1$ of type \texttt{int} to the pair
$\langle\mathtt{int},1\rangle$, which represents the dynamic
value. The corresponding projection checks the type in the first
component against the expected type. It returns the second component
if the types match and throws an exception, otherwise. These
operations can enhance a conventionally typed language
\cite{AbadiCardelliPiercePlotkin1991} or they can form the basis for
optimising a dynamically typed language \cite{Henglein1994}.

Gradual typing is also applicable in a security
context~\cite{DisneyFlanagan2011}. The pure $\lambda$-calculus setting
of the first approach has later been extended to an ML core language.
This extension employs a very liberal treatment of memory references
that are shared between statically and dynamically typed
fragments~\cite{FennellThiemann2013}.

Most security type systems that control information flow and track
data integrity assume an underlying program that is well typed
according to some standard type system~\cite{volpanojcs96}.  Security
labels added as decorations of the standard types indicate the
influence of various peers on the typed value. These labels are mostly
drawn from a lattice of confidentiality or integrity levels, as
already discussed.

For simplicity, existing work on gradual security
typing~\cite{DisneyFlanagan2011,FennellThiemann2013} assumes an
underlying typed program and restricts the gradual aspect of the
system to the security labels.  Gradual security typing guarantees
termination insensitive
noninterference~\cite{GoguenMeseguer1982}. The statically
security-typed parts observe this property by means of the type
system, whereas the dynamically security-typed parts observe the
property with a monitor that enforces the no-sensitive-upgrade policy
on run-time security labels.

Consider the example of a function \texttt{f(s,x)} that manages
information flow with a low-security boolean argument \texttt{s} that
indicates the security level of \texttt{x}, which comes with a dynamic
security level. A gradual security system would annotate the function
as follows.
\begin{verbatim}
f(low s, dyn x) {
  if(!s)
    publish_low(x : dyn => low);
}
\end{verbatim}
The function \texttt{publish\_low} takes a low-security argument and
writes it to a public channel. The function \texttt{f} passes
\texttt{x} to it after applying the coercion from \texttt{dyn} to
\texttt{low} security. This coercion fails if \texttt{s} does not
indicate the security level of \texttt{x} correctly.

In the best case, a coercion from static to dynamic adds run-time
labels whereas a coercion from dynamic to static removes them. While
such a design is possible in the presence of memory references, it
restricts the use of references that are shared between statically and
dynamically typed parts of a program. For that reason, the language
proposed in~\cite{FennellThiemann2013} requires some dynamic checks
even in the statically typed parts of a program.

Subsequent work on gradual annotated types~\cite{FennellThiemann2014}
indicates that the execution model for gradual security with
references can be improved to the point that statically typed parts
need no dynamic checks. Ongoing work considers the formalisation and
implementation of a system improved along these lines in the context
of a Java-like language.

Up to this point, the developments support legacy code (which is
assumed to be typed, but not with a security type system) embedded in
new code, developed with the help of a suitable security type
system. The gradual approach places security coercions at the borders
of the legacy code, potentially adding run-time labels to all values, and
monitors its execution. In contrast, new code would run without labels
at full speed because its security properties are guaranteed by the
static type system.

\subsection{Gradual Security Typing and Sessions}\label{gsts}

Addressing the connection of legacy code with new code via
communication channels is the point where session types or other
behavioural types enter the scene.  The first behavioural type system
that was extended with gradual features was the typestate system of
the concurrent object-oriented language
Plaid~\cite{DBLP:conf/oopsla/SunshineNSAT11,WolffGarciaTanterAldrich2011}. A
typestate system keeps track of the current state of an object
statically. An example is the typestate of a file object that reflects
whether the file is open or closed in its type. As the typestate
changes when operations are applied to the file, it must be linear or
affine, just like the channel type in a session type system. Here is a
very simple file API with typestate:
\begin{align*}
  \mathtt{open} & : \mathtt{filename} \to \mathtt{file}[\cl{open}] \\
  \mathtt{readInt} & : \mathtt{file}[\cl{open}] \to \mathtt{int} \\
  \mathtt{close} & : \mathtt{file}[\cl{open} \leadsto \cl{closed}] \to \mathtt{unit}
\end{align*}
The notation $\cl{open} \leadsto \cl{closed}$ indicates that a file in state $\cl{open}$ is
  expected and that it evolves to state $\cl{closed}$ on executing the function.

The coercion of a value with typestate
into the dynamic type reifies its current typestate in a run-time
value. Operations on the dynamic type step through an automaton that executes the same
transitions as the static typestate computation, similar to communicating
automata~\cite{DBLP:conf/esop/DenielouY12}. The authors of the gradual typestate work suggest to use the
dynamic type during program development because their static typestate
system requires program annotations to manage situations where there
is more than one memory reference to the same object (i.e., aliasing).

The work on gradual types for Plaid cannot be transferred
readily to session types. The obstacle lies in dealing with the
linearity of the channel types, where Plaid resorts to (sophisticated)
alias management. Fortunately, it has been shown that linearity and
gradual typing are largely orthogonal and that many important results from
standard gradual typing carry over 
to a setting with linear types~\cite{FennellThiemann2012-tfp}.  For affine types, it is possible to
gradualise the affine property~\cite{DBLP:conf/esop/TovP10}. It is unlikely that a similar gradualisation
can be achieved for linear types.

One important result in gradual typing is the blame
theorem~\cite{tobindls06}. It strengthens the progress property of a
type system by making precise that failing coercions can always be
attributed to the less precisely typed side of the coercion. In this
context, progress means that a well-typed term is either a value, or
it can perform an evaluation step, or it fails at a cast to a more
precise type. The blame theorem clearly locates the demarcation
between statically proved and dynamically checked code at a particular
kind of coercions that coerce from static to dynamic.

Given these preliminaries, we are now in a position to actually build
a system with gradual session types that also supports verifying
security properties. It is expected that any of the existing static
session systems with security awareness (e.g.,
\cite{capecchiiandc13,CCD15,capecchiconcur10}) can form the basis of a
gradual system, as in the setting without sessions. First steps
towards integrating gradual typing with session types have been taken
\cite{Thiemann2014-tgc}.  

\subsection{Run-time Adaptation}\label{ss:adap}
As (communication-centred) software systems rely on highly dynamic
infrastructures, such as those built on cloud-based platforms,
the ability of adapting to varying requirements and external
conditions becomes crucial to ensure uninterrupted, correct system
behaviour. There is a bidirectional relation between run-time
adaptation and security requirements:
\begin{enumerate}[(a)]
\item On the one hand, it is plausible to react to security threats by
executing an adaptation routine that, e.g., replaces/updates the
affected component;
\item On the other hand, one would like adaptation mechanisms which
address general functional requirements but also preserve
established security policies.  We would like to avoid, e.g.,
mechanisms that update faulty components with correct but insecure
patches.
\end{enumerate}

In the light of this interplay between security and run-time
adaptation, a comprehensive approach that exploits their relationships
appears natural. This is the main motivation of the
paper~\cite{DBLP:journals/corr/CastellaniDP14}, which integrates
security guarantees (access control and secure information flow) with
self-adaptation in a process framework of multiparty structured
communications, exploring the above relation (a). More precisely,
behavioural types with security levels are used to monitor reading and
writing violations, corresponding to access control violation and
information leaks, respectively.  Behavioural types define security
policies by stipulating read and write permissions, represented by
security \emph{levels}.  While a read permission is an upper bound for
the level of incoming messages, a write permission is a lower bound
for the level of outgoing messages.  Accordingly, a reading or writing
violation occurs when a participant attempts to read or write a
message whose level is not allowed by the corresponding read or write
permission. An associated operational semantics is instrumented so as
to trigger adaptation mechanisms in case of violations, but also to
prevent the violations from occurring and propagating their effect in
the choreography.
  
 The framework of~\cite{DBLP:journals/corr/CastellaniDP14} consists of
a language for processes and networks, global types, and run-time
monitors.  Run-time monitors are obtained as projections from global
types onto individual participants.  This way, behavioural types
provide a clear description for enforcing dynamic monitoring of
participants.  Processes represent code that will be coupled with
monitors to implement participants.  A network is a collection of
monitored processes which realise a choreography as described by the
global type.  The semantics of networks includes both \emph{local} and
\emph{global} adaptation mechanisms; their goal is to handle minor and
serious violations, respectively.
 
 Informally, the local adaptation mechanism ``ignores'' unauthorized
actions and modifies a monitored process at run time; it relies on a
\emph{collection} of typed processes, which contains all processes
which may be used in reconfiguration steps.  In case of a read
violation, the local adaptation mechanism modifies the behaviour of
the monitor so as to omit the disallowed read, and then injects a
process from the collection that is compliant with the new monitor. In
case of a write violation, the local adaptation mechanism penalises
the sender by decreasing the read level of his monitor and
replacing the implementation for the receiver.  Here again the new
implementation is extracted from the collection of typed processes.
Therefore, adaptation is local insofar as
reconfiguration steps concern only one monitored process.
   
The global mechanism for adaptation relies on distinguished low-level
values called \emph{nonces}.  When an attempt to leak a value is
detected, a freshly generated nonce is passed instead.  This mechanism
has two goals: first, to avoid improperly communicating the protected
value; second, to allow the whole system to make progress, for the
benefit of the participants not involved in the violation.  At any
point, the semantics may trigger a reconfiguration routine that
replaces the portion of the choreography involving the participants
that may propagate a nonce.  Thus, in the global adaptation mechanism,
a part of the choreography is isolated and replaced, preserving the
correctness of the whole system.  Notice that the function which
returns a new choreography given a choreography with nonces is left
unspecified in~\cite{DBLP:journals/corr/CastellaniDP14}; this function
is intended as a parameter of the operational semantics.
 
To illustrate the kind of scenarios that the framework
in~\cite{DBLP:journals/corr/CastellaniDP14} aims to target (but also
the intuition underlying minor and serious violations), consider a
choreography involving a user, his bank, a store, and a social
network.  Exchanges occur on top of a browser, which relies on
plug-ins to integrate information from different services.  Agreed
exchanges between the user, the bank, and the store may in some cases
lead to a (public) message announcement in the social network.  One
would like to ensure that the buying protocol works as expected, but
also to avoid that sensitive information, exchanged in certain parts
of the protocol, is leaked.  Such an undesired behaviour should be
corrected as soon as possible. In fact, one would like to stop relying
on the (unreliable) participant in ongoing/future instances of the
protocol. Depending on how serious the violation is, however, one may
also like to react in different ways.
\begin{itemize}
\item If the leak is minor (e.g., because the user interacted
  incorrectly with the browser), one may then simply identify the
  source of the leak and postpone the reaction to a later stage,
  enabling unrelated participants in the choreography to proceed with
  their exchanges.
\item Otherwise, if the leak is serious, one may then wish to adapt
  the choreography as soon as possible, removing the plug-in and
  modifying the behaviour of the involved participants.  This form of
  reconfiguration, however, should only concern the participants
  involved with the insecure plug-in; participants not directly
  affected by the leak should not be unnecessarily restarted. In this
  simple example, since the unintended social announcement concerns
  only the user, the store and the social network, updates should not
  affect the behaviour of the bank.
\end{itemize}

Notice that, for simplicity, the framework
in~\cite{DBLP:journals/corr/CastellaniDP14} non-deterministically
selects between local and global mechanisms. This aspect can be
refined, considering that the actual meaning of minor and serious
violations often depends on the applications at hand. Finally, in the
light of the bidirectional relation between adaptation and security
mentioned above, it is clear that the framework
in~\cite{DBLP:journals/corr/CastellaniDP14} follows direction (a)
(i.e., it is a form of security-driven adaptation); addressing
direction (b) (i.e., forms of correctness-driven adaptation that
preserve security policies) is an interesting topic for future work.

\subsection{Role-based Access Control}\label{sec:dynamicdata}

In an open distributed network, it is extremely important to provide
security and protect privacy during transfer and management of
data. In a series of
papers~\cite{dezaniwflp10,dezanitgc06,dezanitcs08,ghilezantar12,jaksicictcs12},
behavioural types for distributed systems containing semi-structured
XML data are investigated.  These works introduce type-based
verification techniques using the model presented
in~\cite{gardnertcs05} (where the focus is on the model, including its
behavioural theory): a network of peers is a parallel composition of
locations, where each location consists of a data tree and a process.
Locations are enriched by security policies prescribing how the access
permissions of roles can be modified.  Processes have roles and edges
in data trees are associated with roles, representing permissions (to
access edges) assigned to roles. Well-typed systems respect prescribed
security policies and role-based access control.

The issues addressed by the above mentioned works, namely role-based
access control~\cite{dezaniwflp10}, are also a main concern when
focussing on communication-centred systems, since the notion of role
may also be involved, in particular in the context of security
protocols. More recently, roles have also gained a behavioural
connotation, as communicating parties may impersonate different roles
throughout their execution and roles may actually be carried out by
several parties, in particular when \emph{delegation} is
involved. Delegation is a challenging feature from a security
perspective since it involves yielding access of a (potentially
secure) medium to third parties.
In~\cite{DBLP:journals/corr/GhilezanJPPV14} a first step in
characterising the delegation of \emph{authorisations} to impersonate
roles is taken, building on the type-based verification techniques
introduced in the approaches mentioned above, in
particular~\cite{dezaniwflp10}, and on the behavioural type system
given in~\cite{baltazartgc12}. Resources are structured, in some
sense, in a way which is similar to that imposed by behavioural
types. Thus, access control for structured resources paves the way to
access control for \emph{communication} resources in a structured
protocol of interaction.



\section{Conclusions} \label{sec:conclusions}

Security and trustworthiness are essential properties for software
systems.  In the context of distributed applications, the challenge of
enforcing these properties is tied to the consistency of structured
conversations among parties.  In fact, since exchanges of (sensitive)
data in such applications often follow predefined communication
sequences, security properties go hand in hand with safety and
liveness properties associated to correct protocols, such as
conformance/com\-pli\-ance, resource usage, and
deadlock-freedom/progress.  As a consequence, the integration of
techniques for describing and enforcing both kinds of properties is
indispensable in many settings. This paper presents an overview of
efforts to achieve this integration in a rigorous way, building upon
calculi and models for communicating processes.  We focus on work
based on \emph{behavioural types}, which extend the well-established
concept of data types to describe complex communication structures.

Our review illustrates how the integration of security concerns into
approaches based on behavioural types leads to a rich landscape of
models and techniques, with both foundational and practical
significance.

On the foundational side, the overview starts with extended models of
session-based communication, which cover a wide variety of
security-related concerns, including access control, secure
information flow, and data integrity.  A fruitful research strand
concerns logic-based approaches to behavioural types.  Approaches
based on linear logic lead to clean, extensible typed models in which
notions of resource-awareness and trustworthy communication have
principled justifications.  In particular, aspects such as
proof-carrying code and digital certificates can be integrated in a
session-typed setting by building upon appropriate (linear) logical
grounds.  From a more practical perspective, we examine ways to
reconcile the usual assumptions of typed models with the actual
requirements of distributed communications over open networks.  These
efforts concern the development of compilers of protocols with
cryptographic information, but also models of honest, contract-based
communication (in which service agreements are handled bottom-up by a
broker), and theories of protection for contracts, which aim at
ensuring honest participants and trusted brokers.  We also discuss
models of gradual typing, in which the combination of static and
dynamic types turns out to be useful to integrate parts of the system
not amenable to static typing (such as legacy code) and to account for
dynamic security policies.  \ma{Behavioural types can be used to
  monitor system execution in case of security violations, and to
  guide adaptations which prevent such violations to occur.}

\begin{table}[t]
\begin{center}
{\small
\begin{tabular}{|c||c|c|c||c|c|}
\multicolumn{1}{c}{}& \multicolumn{3}{c}{Static} & \multicolumn{2}{c}{Dynamic}\\
\hline 
& Enhanced BTs & Contracts & Extended DTs & Gradual & Adaptive\\
\hline
\hline
AC 
 & 
 \begin{tabular}{c} Section~\ref{ss:accon} \\ Section~\ref{sec:dynamicdata}\end{tabular}
 & & Section~\ref{ss:accon} & &
 \\
\hline 
SIF & 
 Section~\ref{ss:secif} & & & Section~\ref{gts} &
 \\
 \hline
Integrity 
 & \begin{tabular}{c} Section~\ref{icd}\\ Section~\ref{sec:proofcarryingcode}\\
 Section~\ref{sec:secure-protocol-compiler}\end{tabular} 
 & 
 Section~\ref{sec:protocolabstraction}
 & Section~\ref{sec:proofcarryingcode}& & Section~\ref{ss:adap}\\
 \hline
Safety & 
 \begin{tabular}{c} Section~\ref{ss:accon}\\Section~\ref{ss:secif}\\Section~\ref{icd}\end{tabular} 
 & & & 
 Section~\ref{gsts}
 &
 \\
\hline
\end{tabular}
}
\end{center}
\caption{Summary of proposals.}
\label{tab:summary}
\end{table}

Table~\ref{tab:summary} summarises the proposals described in this
paper, organised in two dimensions.  The first dimension, organised in
columns, concerns the kind of approach considered: Enhancements of BTs
(Enhanced BTs), Contracts, Extended Datatypes (Extended DTs), Gradual
and Adaptive Types.  The second dimension, represented by rows,
concerns the targeted security properties: Access Control (AC), Secure
Information Flow (SIF), Integrity, and Safety.  It is immediate to see
that enhancing behavioural types (and the model/framework in which
they are considered) is the most common approach to address security
properties.  On a somewhat orthogonal perspective, we may also observe
that several approaches address integrity concerns so as to transport
security properties into untrusted environments.

In our view, strengthening security guarantees via enhanced type
disciplines and the transfer of such guarantees to less controlled
(more realistic) environments constitute two important axes along
which forthcoming research in behavioural types for security should
proceed.  In particular, we believe that the consolidation of
techniques that relate specifications to implementations is a crucial
building block for more reliable and secure applications.  A
considerable body of work has been carried out in the last years on
ensuring ``safe'' protocols by focusing on protocol specifications,
since implementations are too low-level to support reasoning on
protocol properties. By ensuring that implementations conform to
specifications in a rigorous way, the properties established for the
specification may carry over to actual implementations.

For the sort of complex security properties that we have discussed in
this document, we believe that lifting the analysis to the level of
specifications/types enables reasoning at a more adequate abstraction
level. In particular, since it is crucial to consider how such
properties may be transferred to open environments, we need to focus
on techniques that support local reasoning, building upon clearly
defined compositionality principles. Indeed, compositionality is the
key to enable scalable and tractable analyses.

From a practical perspective, an opportunity for future research
concerns integration of models of security monitoring into emerging
practical languages and frameworks based on behavioural types.  For
instance, the Scribble protocol language~\cite{YHNN2013} provides
support for developing large-scale distributed applications whose
interaction architecture can be expressed as multiparty session types.
Scribble has been used for implementing run-time type checking
(monitoring) of communicating
processes~\cite{BocchiCDHY13,ChenBDHY11}, also in collaboration with
the Ocean Observatories Initiative~\cite{OOI}.

\paragraph{{\bf Acknowledgments} We acknowledge support from COST
  Action IC1201 \emph{Behavioural Types for Reliable Large-Scale
    Software Systems} (BETTY) and we thank the members of BETTY
  working group on Security (WG2) for interesting related discussions.  We
  are grateful to the anonymous reviewers for their feedback and many
  insightful remarks, which greatly helped us to improve the quality of
  this document}




\section*{References}

\begin{thebibliography}{10}

\bibitem{AbadiCardelliPiercePlotkin1991}
M.~Abadi, L.~Cardelli, B.~Pierce, and G.~Plotkin.
\newblock Dynamic typing in a statically typed language.
\newblock {\em ACM Trans. Prog. Lang. Syst.}, 13(2):237--268, 1991.

\bibitem{wg3soar}
D.~Ancona, V.~Bono, M.~Bravetti, G.~Castagna, J.~Campos, P.-M. Deni\'elou,
  S.~Gay, N.~Gesbert, E.~Giachino, R.~Hu, E.~B. Johnsen, F.~Martins,
  V.~Mascardi, F.~Montesi, N.~Ng, R.~Neykova, L.~Padovani, V.~Vasconcelos, and
  N.~Yoshida.
\newblock Behavioral types in programming languages.
\newblock
  \url{http://www.behavioural-types.eu/publications/WG3-State-of-the-Art.pdf},
  2014.

\bibitem{baltazartgc12}
P.~Baltazar, L.~Caires, V.~T. Vasconcelos, and H.~T. Vieira.
\newblock A type system for flexible role assignment in multiparty
  communicating systems.
\newblock In {\em TGC 2012}, volume 8191 of {\em LNCS}, pages 82--96. Springer,
  2013.

\bibitem{BMV12}
P.~Baltazar, D.~Mostrous, and V.~T. Vasconcelos.
\newblock Linearly refined session types.
\newblock In {\em {LINEARITY} 2012}, volume 101 of {\em {EPTCS}}, pages 38--49,
  2012.

\bibitem{Barbanera10ppdp}
F.~Barbanera and U.~de'Liguoro.
\newblock Two notions of sub-behaviour for session-based client/server systems.
\newblock In {\em Proc. {PPDP}}, pages 155--164, 2010.

\bibitem{SPARK}
J.~Barnes.
\newblock {\em High Integrity Software: The {SPARK} Approach to Safety and
  Security}.
\newblock Addison-Wesley, 2003.

\bibitem{BCPZ15jlamp}
M.~Bartoletti, T.~Cimoli, G.~M. Pinna, and R.~Zunino.
\newblock Contracts as games on event structures.
\newblock  J. Log. Alg. Meth. Prog. (2015), \url{http://dx.doi.org/10.1016/j.jlamp.
2015.05.001}, in press.

\bibitem{DBLP:conf/post/BartolettiCZ13}
M.~Bartoletti, T.~Cimoli, and R.~Zunino.
\newblock A theory of agreements and protection.
\newblock In {\em POST 2013}, volume 7796 of {\em LNCS}, pages 186--205.
  Springer, 2013.

\bibitem{BMSZ14maude}
M.~Bartoletti, M.~Murgia, A.~Scalas, and R.~Zunino.
\newblock Modelling and verifying contract-oriented systems in {Maude}.
\newblock In {\em {WRLA 2014}}, volume 8663 of {\em LNCS}, pages 130--146.
  Springer, 2014.

\bibitem{BSTZ13forte}
M.~Bartoletti, A.~Scalas, E.~Tuosto, and R.~Zunino.
\newblock Honesty by typing.
\newblock In {\em FMOODS/FORTE 2013}, volume 7892 of {\em LNCS}, pages
  305--320. Springer, 2013.

\bibitem{BTZ12sacs}
M.~Bartoletti, E.~Tuosto, and R.~Zunino.
\newblock Contract-oriented computing in {CO2}.
\newblock {\em Sci. Ann. Comp. Sci.}, 22(1):5--60, 2012.

\bibitem{BTZ12coordination}
M.~Bartoletti, E.~Tuosto, and R.~Zunino.
\newblock On the realizability of contracts in dishonest systems.
\newblock In {\em COORDINATION 2012}, volume 7274 of {\em LNCS}, pages
  245--260. Springer, 2012.

\bibitem{BCDDDY08}
L.~Bettini, M.~Coppo, L.~D'Antoni, M.~D. Luca, M.~Dezani{-}Ciancaglini, and
  N.~Yoshida.
\newblock Global progress in dynamically interleaved multiparty sessions.
\newblock In {\em CONCUR 2008}, volume 5201 of {\em LNCS}, pages 418--433.
  Springer, 2008.

\bibitem{bhargavancsf09}
K.~Bhargavan, R.~Corin, P.-M. Deni{\'e}lou, C.~Fournet, and J.~J. Leifer.
\newblock Cryptographic protocol synthesis and verification for multiparty
  sessions.
\newblock In {\em CSF 2009}, pages 124--140. IEEE, 2009.

\bibitem{BocchiCDHY13}
L.~Bocchi, T.~Chen, R.~Demangeon, K.~Honda, and N.~Yoshida.
\newblock Monitoring networks through multiparty session types.
\newblock In {\em FMOODS/FORTE 2013}, volume 7892 of {\em LNCS}, pages 50--65.
  Springer, 2013.

\bibitem{DBLP:conf/concur/BocchiHTY10}
L.~Bocchi, K.~Honda, E.~Tuosto, and N.~Yoshida.
\newblock A theory of design-by-contract for distributed multiparty
  interactions.
\newblock In {\em CONCUR 2010}, volume 6269 of {\em LNCS}, pages 162--176.
  Springer, 2010.

\bibitem{bonellijfp05}
E.~Bonelli, A.~B. Compagnoni, and E.~L. Gunter.
\newblock Correspondence assertions for process synchronization in concurrent
  communications.
\newblock {\em J. Funct. Program.}, 15(2):219--247, 2005.

\bibitem{BBCNLLMMRSVZ06}
M.~Boreale, R.~Bruni, L.~Caires, R.~{De Nicola}, I.~Lanese, M.~Loreti,
  F.~Martins, U.~Montanari, A.~Ravara, D.~Sangiorgi, V.~T. Vasconcelos, and
  G.~Zavattaro.
\newblock {SCC}: A service centered calculus.
\newblock In {\em WS-FM 2006}, volume 4184 of {\em LNCS}, pages 38--57.
  Springer, 2006.

\bibitem{Bossi-Focardi-Piazza-Rossi'04}
A.~Bossi, R.~Focardi, C.~Piazza, and S.~Rossi.
\newblock Verifying persistent security properties.
\newblock {\em Comp. Lang., Syst. {\&} Struct.}, 30(3-4):231--258, 2004.

\bibitem{BM08}
R.~Bruni and L.~G. Mezzina.
\newblock Types and deadlock freedom in a calculus of services, sessions and
  pipelines.
\newblock In {\em AMAST 2008}, volume 5140 of {\em LNCS}, pages 100--115.
  Springer, 2008.

\bibitem{DBLP:conf/concur/CairesP10}
L.~Caires and F.~Pfenning.
\newblock Session types as intuitionistic linear propositions.
\newblock In {\em CONCUR 2010}, volume 6269 of {\em LNCS}, pages 222--236.
  Springer, 2010.

\bibitem{capecchiiandc13}
S.~Capecchi, I.~Castellani, and M.~Dezani-Ciancaglini.
\newblock Typing access control and secure information flow in sessions.
\newblock {\em Inf. Comput.}, 238:68--105, 2014.

\bibitem{CCD15}
S.~Capecchi, I.~Castellani, and M.~Dezani-Ciancaglini.
\newblock Information flow safety in multiparty sessions.
\newblock {\em Math. Struct. in Comp. Science} (2015),
\newblock \url{http://journals.cambridge.org/abstract_S0960129514000619}.

\bibitem{capecchiconcur10}
S.~Capecchi, I.~Castellani, M.~Dezani-Ciancaglini, and T.~Rezk.
\newblock Session types for access and information flow control.
\newblock In {\em CONCUR 2010}, volume 6269 of {\em LNCS}, pages 237--252.
  Springer, 2010.

\bibitem{DBLP:journals/corr/CastellaniDP14}
I.~Castellani, M.~Dezani{-}Ciancaglini, and J.~A. P{\'{e}}rez.
\newblock Self-adaptation and secure information flow in multiparty structured
  communications: {A} unified perspective.
\newblock In {\em BEAT 2014}, volume 162 of {\em {EPTCS}}, pages 9--18, 2014.

\bibitem{ChenBDHY11}
T.-C. Chen, L.~Bocchi, P.-M. Deni{\'e}lou, K.~Honda, and N.~Yoshida.
\newblock Asynchronous distributed monitoring for multiparty session
  enforcement.
\newblock In {\em TGC 2011}, volume 7173 of {\em LNCS}, pages 25--45. Springer,
  2012.

\bibitem{jif}
S.~Chong, A.~C. Myers, K.~Vikram, and L.~Zheng.
\newblock {\em Jif Reference Manual}.
\newblock Cornell University, 2009.
\newblock \url{http://www.cs.cornell.edu/jif}.

\bibitem{corinD07}
R.~Corin and P.-M. Deni{\'e}lou.
\newblock A protocol compiler for secure sessions in {ML}.
\newblock In {\em TGC 2007}, volume 4912 of {\em LNCS}, pages 276--293.
  Springer, 2007.

\bibitem{corincsf07}
R.~Corin, P.-M. Deni{\'e}lou, C.~Fournet, K.~Bhargavan, and J.~J. Leifer.
\newblock Secure implementations for typed session abstractions.
\newblock In {\em CSF 2007}, pages 170--186. IEEE, 2007.

\bibitem{corinjcs08}
R.~Corin, P.-M. Deni{\'e}lou, C.~Fournet, K.~Bhargavan, and J.~J. Leifer.
\newblock A secure compiler for session abstractions.
\newblock {\em J. Comp. Sec.}, 16(5):573--636, 2008.

\bibitem{Crafa-Rossi'05}
S.~Crafa and S.~Rossi.
\newblock A theory of noninterference for the pi-calculus.
\newblock In {\em TGC 2005}, volume 3705 of {\em LNCS}, pages 2--18. Springer,
  2005.

\bibitem{DBLP:conf/esop/DenielouY12}
P.~Deni{\'{e}}lou and N.~Yoshida.
\newblock Multiparty session types meet communicating automata.
\newblock In {\em ESOP 2012}, volume 7211 of {\em LNCS}, pages 194--213.
  Springer, 2012.

\bibitem{DenningD77}
D.~E. Denning and P.~J. Denning.
\newblock Certification of programs for secure information flow.
\newblock {\em Commun. {ACM}}, 20(7):504--513, 1977.

\bibitem{DBLP:journals/iandc/Dezani-CiancagliniDMY09}
M.~Dezani{-}Ciancaglini, S.~Drossopoulou, D.~Mostrous, and N.~Yoshida.
\newblock Objects and session types.
\newblock {\em Inf. Comput.}, 207(5):595--641, 2009.

\bibitem{dezaniwflp10}
M.~Dezani-Ciancaglini, S.~Ghilezan, S.~Jaksic, and J.~Pantovic.
\newblock Types for role-based access control of dynamic web data.
\newblock In {\em WFLP 2010}, volume 6559 of {\em LNCS}, pages 1--29. Springer,
  2010.

\bibitem{dezanitgc06}
M.~Dezani-Ciancaglini, S.~Ghilezan, and J.~Pantovic.
\newblock Security types for dynamic web data.
\newblock In {\em TGC 2006}, volume 4661 of {\em LNCS}, pages 263--280.
  Springer, 2006.

\bibitem{dezanitcs08}
M.~Dezani-Ciancaglini, S.~Ghilezan, J.~Pantovic, and D.~Varacca.
\newblock Security types for dynamic web data.
\newblock {\em Theor. Comput. Sci.}, 402(2-3):156--171, 2008.

\bibitem{DisneyFlanagan2011}
T.~Disney and C.~Flanagan.
\newblock Gradual information flow typing.
\newblock In {\em STOP 2011}, 2011.

\bibitem{DBLP:conf/eurosys/FahndrichAHHHLL06}
M.~F{\"{a}}hndrich, M.~Aiken, C.~Hawblitzel, O.~Hodson, G.~C. Hunt, J.~R.
  Larus, and S.~Levi.
\newblock Language support for fast and reliable message-based communication in
  singularity {OS}.
\newblock In {\em EuroSys 2006}, pages 177--190. {ACM}, 2006.

\bibitem{FennellThiemann2012-tfp}
L.~Fennell and P.~Thiemann.
\newblock The blame theorem for a linear lambda calculus with type dynamic.
\newblock In {\em {TFP} 2012}, volume 7829 of {\em LNCS}, pages 37--52.
  Springer, 2012.

\bibitem{FennellThiemann2013}
L.~Fennell and P.~Thiemann.
\newblock Gradual security typing with references.
\newblock In {\em CSF 2013}, pages 224--239. IEEE, 2013.

\bibitem{FennellThiemann2014}
L.~Fennell and P.~Thiemann.
\newblock Gradual typing for annotated type systems.
\newblock In {\em ESOP 2014}, volume 8410 of {\em LNCS}, pages 47--66.
  Springer, 2014.

\bibitem{Focardi-Gorrieri'01}
R.~Focardi and R.~Gorrieri, editors.
\newblock {\em Foundations of Security Analysis and Design, Tutorial Lectures},
  volume 2171 of {\em LNCS}. Springer, 2001.

\bibitem{gardnertcs05}
P.~Gardner and S.~Maffeis.
\newblock Modelling dynamic web data.
\newblock {\em Theor. Comput. Sci.}, 342(1):104--131, 2005.

\bibitem{Garg10oakland}
D.~Garg and F.~Pfenning.
\newblock A proof-carrying file system.
\newblock In {\em S{\&}P 2010}, pages 349--364. IEEE, 2010.

\bibitem{DBLP:journals/acta/GayH05}
S.~J. Gay and M.~Hole.
\newblock Subtyping for session types in the pi calculus.
\newblock {\em Acta Inf.}, 42(2-3):191--225, 2005.

\bibitem{ghilezantar12}
S.~Ghilezan, S.~Jaksic, J.~Pantovic, and M.~Dezani-Ciancaglini.
\newblock Types and roles for web security.
\newblock {\em Trans. Adv. Res.}, 8(2):16--21, 2012.

\bibitem{DBLP:journals/corr/GhilezanJPPV14}
S.~Ghilezan, S.~Jaksic, J.~Pantovic, J.~A. P{\'{e}}rez, and H.~T. Vieira.
\newblock Dynamic role authorization in multiparty conversations.
\newblock In {\em {BEAT} 2014}, volume 162 of {\em {EPTCS}}, pages 1--8, 2014.

\bibitem{Girard87}
J.-Y. Girard.
\newblock {Linear Logic}.
\newblock {\em Theor. Comput. Sci.}, 50:1--102, 1987.

\bibitem{GoguenMeseguer1982}
J.~A. Goguen and J.~Meseguer.
\newblock Security policies and security models.
\newblock In {\em IEEE Symp. Sec. and Priv.}, pages 11--20, 1982.

\bibitem{GJ03}
A.~D. Gordon and A.~Jeffrey.
\newblock Typing correspondence assertions for communication protocols.
\newblock {\em Theor. Comput. Sci.}, 300(1-3):379--409, 2003.

\bibitem{Harper:1993:FDL:138027.138060}
R.~Harper, F.~Honsell, and G.~Plotkin.
\newblock A framework for defining logics.
\newblock {\em J. ACM}, 40:143--184, 1993.

\bibitem{HedinBBS14}
D.~Hedin, A.~Birgisson, L.~Bello, and A.~Sabelfeld.
\newblock {JSFlow}: tracking information flow in {JavaScript} and its {API}s.
\newblock In {\em {SAC} 2014}, pages 1663--1671. {ACM}, 2014.

\bibitem{Henglein1994}
F.~Henglein.
\newblock Dynamic typing: Syntax and proof theory.
\newblock {\em Sci. Comput. Programming}, 22:197--230, 1994.

\bibitem{Hennessy'05}
M.~Hennessy.
\newblock The security pi-calculus and non-interference.
\newblock {\em J. Log. Algebr. Program.}, 63(1):3--34, 2005.

\bibitem{Hennessy-Riely'02}
M.~Hennessy and J.~Riely.
\newblock Information flow vs. resource access in the asynchronous pi-calculus.
\newblock {\em {ACM} Trans. Program. Lang. Syst.}, 24(5):566--591, 2002.

\bibitem{DBLP:conf/concur/Honda93}
K.~Honda.
\newblock Types for dyadic interaction.
\newblock In {\em CONCUR 1993}, volume 715 of {\em LNCS}, pages 509--523.
  Springer, 1993.

\bibitem{DBLP:conf/esop/HondaVK98}
K.~Honda, V.~T. Vasconcelos, and M.~Kubo.
\newblock Language primitives and type discipline for structured
  communication-based programming.
\newblock In {\em ESOP 1998}, volume 1381 of {\em LNCS}, pages 122--138.
  Springer, 1998.

\bibitem{Honda-Vasc-el'00}
K.~Honda, V.~T. Vasconcelos, and N.~Yoshida.
\newblock Secure information flow as typed process behaviour.
\newblock In {\em ESOP 2000}, volume 1782 of {\em LNCS}, pages 180--199.
  Springer, 2000.

\bibitem{Honda-Yoshida'07}
K.~Honda and N.~Yoshida.
\newblock A uniform type structure for secure information flow.
\newblock {\em {ACM} Trans. Program. Lang. Syst.}, 29(6), 2007.

\bibitem{CHY08}
K.~Honda, N.~Yoshida, and M.~Carbone.
\newblock {Multiparty Asynchronous Session Types}.
\newblock In {\em POPL 2008}, pages 273--284. ACM, 2008.

\bibitem{wg1soar}
H.~H{\"{u}}ttel, I.~Lanese, V.~T. Vasconcelos, L.~Caires, M.~Carbone,
  P.~Deni{\'{e}}lou, D.~Mostrous, L.~Padovani, A.~Ravara, E.~Tuosto, H.~T.
  Vieira, and G.~Zavattaro.
\newblock Foundations of behavioural types.
\newblock
  \url{http://www.behavioural-types.eu/publications/WG1-State-of-the-Art.pdf},
  2014.

\bibitem{jaksicictcs12}
S.~Jaksic.
\newblock Input/output types for dynamic web data.
\newblock In {\em ICTCS 2012}, 2012.

\bibitem{Orc2009}
D.~Kitchin, A.~Quark, W.~R. Cook, and J.~Misra.
\newblock The {Orc} programming language.
\newblock In {\em FMOODS/FORTE 2009}, volume 5522 of {\em LNCS}, pages 1--25.
  Springer, 2009.

\bibitem{K02}
N.~Kobayashi.
\newblock {A Type System for Lock-Free Processes}.
\newblock {\em Inf. Comput.}, 177:122--159, 2002.

\bibitem{Kobayashi'05}
N.~Kobayashi.
\newblock Type-based information flow analysis for the pi-calculus.
\newblock {\em Acta Inf.}, 42(4-5):291--347, 2005.

\bibitem{kolundzijawsfm08}
M.~Kolundzija.
\newblock Security types for sessions and pipelines.
\newblock In {\em WS-FM 2008}, volume 5387 of {\em LNCS}, pages 175--190.
  Springer, 2008.

\bibitem{lapadulafsen07}
A.~Lapadula, R.~Pugliese, and F.~Tiezzi.
\newblock Regulating data exchange in service oriented applications.
\newblock In {\em FSEN 2007}, volume 4767 of {\em LNCS}, pages 223--239.
  Springer, 2007.

\bibitem{Mukhija2007qos}
A.~Mukhija, A.~Dingwall-Smith, and D.~Rosenblum.
\newblock {QoS}-aware service composition in {Dino}.
\newblock In {\em {ECOWS}}, pages 3--12, 2007.

\bibitem{neculapopl97}
G.~C. Necula.
\newblock Proof-carrying code.
\newblock In {\em POPL 1997}, pages 106--119. ACM, 1997.

\bibitem{OOI}
{{O}cean {O}bservatories {I}nitiative}, 2010.
\newblock
  \url{http://www.oceanleadership.org/programs-and-partnerships/ocean-observing/ooi/}.

\bibitem{P13}
L.~Padovani.
\newblock Deadlock and lock freedom in the linear {\(\pi\)}-calculus.
\newblock In {\em CSL-LICS 2014}, page~72. {ACM}, 2014.

\bibitem{Perez2014}
J.~A. P{\'{e}}rez, L.~Caires, F.~Pfenning, and B.~Toninho.
\newblock Linear logical relations and observational equivalences for
  session-based concurrency.
\newblock {\em Inf. Comput.}, 239:254--302, 2014.

\bibitem{Pfenning:2001:IEP:871816.871845}
F.~Pfenning.
\newblock Intensionality, extensionality, and proof irrelevance in modal type
  theory.
\newblock In {\em LICS 2001}, pages 221--230. IEEE, 2001.

\bibitem{DBLP:conf/cpp/PfenningCT11}
F.~Pfenning, L.~Caires, and B.~Toninho.
\newblock Proof-carrying code in a session-typed process calculus.
\newblock In {\em CPP 2011}, volume 7086 of {\em LNCS}, pages 21--36. Springer,
  2011.

\bibitem{planulconcur09}
J.~Planul, R.~Corin, and C.~Fournet.
\newblock Secure enforcement for global process specifications.
\newblock In {\em CONCUR 2009}, volume 5710 of {\em LNCS}, pages 511--526.
  Springer, 2009.

\bibitem{Pottier'02}
F.~Pottier.
\newblock A simple view of type-secure information flow in the p-calculus.
\newblock In {\em CSFW 2002}, pages 320--330. IEEE, 2002.

\bibitem{PSS05}
F.~Pottier, C.~Skalka, and S.~F. Smith.
\newblock A systematic approach to static access control.
\newblock {\em ACM Trans. Program. Lang. Syst.}, 27(2):344--382, 2005.

\bibitem{PT12}
R.~Pugliese and F.~Tiezzi.
\newblock A calculus for orchestration of web services.
\newblock {\em J. Applied Logic}, 10(1):2--31, 2012.

\bibitem{Ryan-Schneider'99}
P.~Y.~A. Ryan and S.~A. Schneider.
\newblock Process algebra and non-interference.
\newblock In {\em CSFW 1999}, pages 214--227. {IEEE}, 1999.

\bibitem{SantosR14}
J.~F. Santos and T.~Rezk.
\newblock An information flow monitor-inlining compiler for securing a core of
  {JavaScript}.
\newblock In {\em SEC 2014}, volume 428 of {\em IFIP Adv. ICT}, pages 278--292.
  Springer, 2014.

\bibitem{SiekTaha2007}
J.~Siek and W.~Taha.
\newblock Gradual typing for objects.
\newblock In {\em ECOOP 2007}, volume 4609 of {\em LNCS}, pages 2--27.
  Springer, 2007.

\bibitem{SiekTaha2006}
J.~G. Siek and W.~Taha.
\newblock Gradual typing for functional languages.
\newblock In {\em Scheme and Functional Programming Workshop}, pages 81--92,
  2006.

\bibitem{flowcaml}
V.~Simonet.
\newblock {\em The Flow Caml System: Documentation and userÕs manual}.
\newblock INRIA, 2003.
\newblock \url{http://www.normalesup.org/~simonet/soft/flowcaml/}.

\bibitem{DBLP:conf/oopsla/SunshineNSAT11}
J.~Sunshine, K.~Naden, S.~Stork, J.~Aldrich, and {\'E}.~Tanter.
\newblock First-class state change in {Plaid}.
\newblock In {\em OOPSLA 2011}, pages 713--732. ACM, 2011.

\bibitem{DBLP:journals/jfp/SwamyCFSBY13}
N.~Swamy, J.~Chen, C.~Fournet, P.~Strub, K.~Bhargavan, and J.~Yang.
\newblock Secure distributed programming with value-dependent types.
\newblock {\em J. Funct. Program.}, 23(4):402--451, 2013.

\bibitem{Thiemann2014-tgc}
P.~Thiemann.
\newblock Gradual typing for session types.
\newblock In {\em TGC 2014}, volume 8902 of {\em LNCS}, pages 144--158.
  Springer, 2014.

\bibitem{tobindls06}
S.~Tobin{-}Hochstadt and M.~Felleisen.
\newblock Interlanguage migration: from scripts to programs.
\newblock In {\em OOPSLA 2006}, pages 964--974. {ACM}, 2006.

\bibitem{toninho11ppdp}
B.~Toninho, L.~Caires, and F.~Pfenning.
\newblock Dependent session types via intuitionistic linear type theory.
\newblock In {\em PPDP 2011}, pages 161--172. ACM, 2011.

\bibitem{DBLP:conf/esop/TovP10}
J.~A. Tov and R.~Pucella.
\newblock Stateful contracts for affine types.
\newblock In {\em ESOP 2010}, volume 6012 of {\em LNCS}, pages 550--569.
  Springer, 2010.

\bibitem{DBLP:journals/fuin/VallecilloVR06}
A.~Vallecillo, V.~T. Vasconcelos, and A.~Ravara.
\newblock Typing the behavior of software components using session types.
\newblock {\em Fundam. Inform.}, 73(4):583--598, 2006.

\bibitem{DBLP:journals/tcs/VasconcelosGR06}
V.~T. Vasconcelos, S.~J. Gay, and A.~Ravara.
\newblock Type checking a multithreaded functional language with session types.
\newblock {\em Theor. Comput. Sci.}, 368(1-2):64--87, 2006.

\bibitem{volpanojcs96}
D.~Volpano, C.~Irvine, and G.~Smith.
\newblock {A Sound Type System for Secure Flow Analysis}.
\newblock {\em J. Comp. Sec.}, 4(2-3):167--187, 1996.

\bibitem{DBLP:conf/ac/Winskel86}
G.~Winskel.
\newblock Event structures.
\newblock In {\em APN 1986}, volume 255 of {\em LNCS}, pages 325--392.
  Springer, 1986.

\bibitem{WolffGarciaTanterAldrich2011}
R.~Wolff, R.~Garcia, {\'E}.~Tanter, and J.~Aldrich.
\newblock Gradual typestate.
\newblock In {\em ECOOP 2011}, volume 6813 of {\em LNCS}, pages 459--483.
  Springer, 2011.

\bibitem{YHNN2013}
N.~Yoshida, R.~Hu, R.~Neykova, and N.~Ng.
\newblock The {S}cribble protocol language.
\newblock In {\em TGC 2013}, volume 8358 of {\em LNCS}, pages 22--41. Springer,
  2013.

\end{thebibliography}
\end{document}